\documentclass[english]{revtex4}
\usepackage[T1]{fontenc}
\usepackage[latin9]{inputenc}
\usepackage{amsmath}
\usepackage{amssymb}
\usepackage{graphicx}
\usepackage{wasysym}

\makeatletter
\@ifundefined{textcolor}{}
{%
 \definecolor{BLACK}{gray}{0}
 \definecolor{WHITE}{gray}{1}
 \definecolor{RED}{rgb}{1,0,0}
 \definecolor{GREEN}{rgb}{0,1,0}
 \definecolor{BLUE}{rgb}{0,0,1}
 \definecolor{CYAN}{cmyk}{1,0,0,0}
 \definecolor{MAGENTA}{cmyk}{0,1,0,0}
 \definecolor{YELLOW}{cmyk}{0,0,1,0}
}

\makeatother

\usepackage{babel}
\begin{document}
\title{Quantum Modified Gravity at Low Energy in the Ricci Flow of Quantum
Spacetime}
\author{M.J.Luo}
\address{Department of Physics, Jiangsu University, Zhenjiang 212013, People's
Republic of China}
\email{mjluo@ujs.edu.cn}

\begin{abstract}
Quantum treatment of physical reference frame leads to the Ricci flow
of quantum spacetime, which is a quite rigid framework to quantum
and renormalization effect of gravity. The theory has a low characteristic
energy scale described by a unique constant: the critical density
of the universe. At low energy long distance (cosmic or galactic)
scale, the theory modifies Einstein's gravity which naturally gives
rise to a cosmological constant as a counter term of the Ricci flow
at leading order and an effective scale dependent Einstein-Hilbert
action. 

In the weak and static gravity limit, the framework gives rise to
a transition trend away from Newtonian gravity and similar to the
MOdified Newtonian Dynamics (MOND) around the characteristic scale.
When local curvature is large, Newtonian gravity is recovered. When
local curvature is low enough to be comparable with the asymptotic
background curvature corresponding to the characteristic energy scale,
the transition trend produces the baryonic Tully-Fisher relation.
For intermediate general curvature around the background curvature,
the interpolating Lagrangian function yields a similar transition
trend to the observed radial acceleration relation of galaxies. When
the baryonic matter density is much lower than the critical density
at the outskirt of a galaxy, there may be a universal \textquotedblleft acceleration
floor\textquotedblright{} corresponding to the acceleration expansion
of the universe, which differs from MOND at its deep-MOND limit. 

The critical acceleration constant $a_{0}$ introduced in MOND is
related to the low characteristic energy scale of the theory. The
cosmological constant gives a universal leading order contribution
to $a_{0}$ and the flow effect gives the next order scale dependent
contribution, which equivalently induces the ``cold dark matter''
to the theory. $a_{0}$ is consistent with galaxian data when the
``dark matter'' is about 5 times the baryonic matter.
\end{abstract}
\maketitle

\section{Introduction}

A wealth of astronomical observations indicate the presence of missing
masses or acceleration discrepancies in the universe based on the
classical gravity theory (general relativity) although the theory
is well tested within solar system very precisely. One possible approach
to solve the problem is by separately introducing the missing masses
components into the universe, for instance, the dark energy (DE or
the cosmological constant $\Lambda$ (CC) ) (Equation Of State $w=-1$)
and cold dark matter (CDM) ($w\approx0$) in the so called $\Lambda$CDM-model.
Another approach is by modifying the law of gravity, within which
the problem should be more appropriately reconsidered as a gravity/acceleration
discrepancy between the (cosmic or galactic) long distance scale and
the (solar system or laboratory) short distance scale. There are some
phenomenological supports for the latter approach, since both the
acceleration expansion of universe (corresponding to the DE or $\Lambda$)
and galactic rotation/acceleration anomalies (corresponding to the
CDM) empirically manifest a particular acceleration scale $a_{0}\approx1.2\times10^{-10}m/s^{2}\approx\frac{\sqrt{\Lambda}}{\left(6\sim8\right)}$,
first proposed in the MOdified Newtonian Dynamics (MOND) by Milgrom
\citep{1983ApJ} (see reviews \citep{Famaey:2011kh,Milgrom:2014usa}
and references therein, or long publication list of Milgrom's). The
baryonic Tully-Fisher law \citep{1977Reprint,2000The} and an amazing
``mass discrepancy-acceleration relation'' \citep{Stacy2016Radial}
with little scatter are also observed, which do not occur naturally
in the $\Lambda$CDM-model. Although the modified gravity approach
might face its own difficulties (e.g. MOND without CDM is failed in
fitting the third and subsequent acoustic peaks in the Cosmic Microwave
Background (CMB)), this line of thinking might lead us to a more ambitious
and unified view to our universe. The internal relation between the
cosmological constant and MOND has been generally conjectured, and
varieties of underpinning proposals and possible relativistic generalizations
of MOND are suggested in literature, they are still more or less similar
with the Kepler's law as a phenomenological description, there is
no first principle to determine the exact form of the interpolating
function between the standard gravity limit and the modified one,
thus lacking a fundamental underlying principle and theoretical framework
remains its essential weakness.

Recent years the author based on the quantum treatment of physical
reference frame, proposed a framework of quantum spacetime and gravity
\citep{Luo2014The,Luo2015Dark,Luo:2015pca,Luo:2019iby,Luo:2021zpi,Luo:2022goc,Luo:2022statistics,Luo:2021wdh}.
The basic idea of the theory is that when quantum theory is reformulated
on the new foundation of relational quantum state (an entangled state)
describing the ``relation'' between a state of a under-studied quantum
system and a state of a quantum spacetime reference frame system,
a gravitational theory is automatically contained in the quantum framework.
Gravitational phenomenon is given by a relational quantum state describing
a relative motion of the under-studied quantum system with respect
to the material quantum reference frame system. And the 2nd order
central moment of quantum fluctuations of the quantum reference system
introduces the Ricci flow to the quantum spacetime,
\begin{equation}
\frac{\partial g_{\mu\nu}}{\partial t}=-2R_{\mu\nu},\label{eq:ricci flow}
\end{equation}
where $g_{\mu\nu}$ and $R_{\mu\nu}$ are the metric and Ricci curvature
of spacetime, and $t$ is the flow parameter. 

The Ricci flow is historically invented independently from the physics
and mathematics points of views. From the physics angle, the Ricci
flow was introduced by Friedan \citep{friedan1980nonlinear,Friedan1980}
as a renormalization flow of a non-linear sigma model in $2+\epsilon$
dimension. From the mathematics angle, the Ricci flow was introduced
first by Hamilton \citep{Hamilton1982Three} as a useful tool to deform
an initial Riemannian manifold into a more and more ``simple'' and
``good'' manifold whose topology is conserved and finally can be
easily recognized in order to proof geometric theorems (like the Poincare
conjecture). But certain singularity may develop during the Ricci
flow and becomes the stumbling block of Hamilton's program. Around
2003, Perelman introduced several monotonic functionals \citep{perelman2002entropy,perelman2003ricci,perelman307245finite}
to control the singularity during the Ricci flow. What Perelman treated
is in fact a density manifold $(M^{D},g,u)$ with density $u$ as
a generalization of the Riemanian manifold $(M^{D},g)$, in which
the density $u$ describes a local density of the Riemannian manifold
and physically coming from the quantum fluctuation or uncertainty
at each point of the manifold. The Ricci-DeTurck flow 
\begin{equation}
\frac{\partial g_{\mu\nu}}{\partial t}=-2\left(R_{\mu\nu}-\nabla_{\mu}\nabla_{\nu}\log u\right)\label{eq:ricci-deturck}
\end{equation}
of the density manifold (equivalent to the Ricci flow (\ref{eq:ricci flow})
up to a diffeomorphism given by the gradient of the $u$ density)
is shown to be the gradient flow of Perelman's functionals, so that
he could overcome the stumbling block by using his functionals and
finally complete the Hamilton's program.

In fact the underlying physics of Perelman's formalism is not fully
clear for physicist, the quantum spacetime reference frame picture
is proposed by the author to lay the physical foundation. In the framework
of the quantum spacetime reference frame, the spacetime is measured
by physical rods and clocks as reference frame system and hence subject
to quantum fluctuation. When the quantum fluctuation of the reference
frame system is unimportant (mean fields approximation), the quantum
framework recovers the standard textbook quantum theory without gravity.
When the 2nd order central moment quantum fluctuation as the quantum
correction of the reference frame system is important and be taken
into account (Gaussian approximation), Ricci flow and gravity emerge
in the quantum framework, as if one introduces gravitation into the
standard textbook quantum mechanics. The physical reference frame
modeled by the frame fields system is prepared and calibrated in a
laboratory, which is mathematically described by a non-linear sigma
model using lab's spacetime dimension $4-\epsilon$ (a more practical
example is a multi-wire chamber using the electrons as frame fields
to measure the coordinates of events in a laboratory). An under-studied
quantum system (e.g. the events) has physical meaning only with respect
to the quantum reference frame system (the multi-wire chamber). The
2nd order central moments of quantum fluctuations of the reference
frame fields blur the event and equivalently give quantum variance
to the spacetime coordinates. The variance of the coordinates directly
modifies the quadratic form metric of the Riemannian spacetime geometry,
making the spacetime vary with the scale of the quantum fluctuation.
Such scale dependent quantum correction to the metric continuously
deforms the spacetime geometry driven by its Ricci curvature, which
is exactly the Ricci flow: a renormalization flow of the spacetime.
The $t$ parameter is related to the cutoff energy scale of the Fourier
components of the spacetime coordinates promoted to be quantum frame
fields. As the Ricci flow starts from short distance scale (UV) $t\rightarrow-\infty$
and flows to long distance scale (IR) $t\rightarrow0$, or from the
astronomical viewpoint, the energies of spectral lines (as the tracers
of astronomical observations) start from short distance laboratory
scale and are redshifted to long distance galactic or cosmic scale.
During the process, the spacetime coordinates and metric at a long
distance scale $t$ are given by averaged out the shorter distance
finer details which produces an effective correction to them. In a
more intuitive picture, as the wave pocket of the reference frame
fields, such as the spectral lines, gradually Gaussian (2nd order)
broaden when they travel a long distance, at long distance (e.g. cosmic
or galactic) scale, the 2nd moment (i.e. the intrinsic spectral lines
broadening) correction to the spacetime coordinate or metric becomes
significant and hence can not be ignored. The 2nd order moment quantum
fluctuation of spacetime gives rise to correction to the 2nd or quadratic
order, thus quantities like curvature or acceleration as the second
spacetime derivative obtain additional coarse-graining corrections
in a natural and rigid way at long distance scale, which is considered
as the root of the acceleration discrepancies in astronomical scale
observations and the quantum modified gravity at low energy. 

Further, in the framework, the 2nd order quantum corrections to gravity
and acceleration are in a universal way, so that the correction is
not merely the correction to specific spectral line itself but the
correction to the spacetime. In this sense the Equivalence Principle
retains at the quantum level, which lays the physical foundation for
the physical measurement of spacetime geometry and geometric description
of gravity. The quantum description of the spacetime reference frame
together with the quantum version of the Equivalence Principle leads
to a completely different view on the behavior of quantum modified
gravity: it is at the long distance scale where the quantum correction
is significant. Beside the above intuitive ``wave pocket broaden''
picture of spacetime fuzziness at long distance scale, it is also
reflected in the characteristic scale of the gravity theory described
by the only input dimensional constant $\lambda$ of the quantum spacetime
reference frame (in fact the only input constant for the $d=4-\epsilon$
non-linear sigma model). As we will see in the section-II-B that,
to recover the standard Einstein gravity the constant must be exactly
the critical density $\lambda=\frac{3H_{0}^{2}}{8\pi G}=\frac{\Lambda}{\Omega_{\Lambda}8\pi G}\approx\left(10^{-3}\mathrm{eV}\right)^{4}$.
In contrast to the general believing of quantum gravity, the input
constant is not the single Newton's constant but the critical density
$\lambda$ of the universe, as a combination of the Newton's constant
$G$ and Hubble's constant $H_{0}$. As a consequence, the characteristic
energy scale of the gravity theory is not the Planck scale, but the
low critical density scale which is a long distance cosmic scale.
As the $t$ parameter in the framework is a ratio of the cutoff energy
scale $k^{2}$ of the frame fields over the critical density, $t=-\frac{1}{64\pi^{2}\lambda}k^{2}$,
when the energy scale of the frame fields is highly redshifted by
the scale factor $\mathbf{a}^{2}$, i.e. $k^{2}\propto\mathbf{a}^{2}\rightarrow0$,
or $t\rightarrow0^{-}$, it is at the low energy limit that gravity
is strongly modified. In this paper, when we mention ``scale $t$''
(or later $\tau$) of an astronomical object, it can be understood
physically as the the scale factor $t\propto-\mathbf{a}^{2}$ or related
redshift in the sense of the standard expanding universe picture. 

An important feature of the framework is that the critical density
$\lambda$ is the characteristic scale of the quantum gravity, as
a consequence, the cosmological constant problem appearing in the
naive quantum general relativity is more readily understood. Since
the natural scale of the cosmological constant is no longer the Planck
scale, which is $10^{120}$ times the observed value, but of order
of the critical density $\lambda$. And the fraction in the critical
density $\Omega_{\Lambda}=\frac{\Lambda}{8\pi G\lambda}\approx0.7$
of order one is given by the counter term to the spacetime volume
flow, which is related to a Ricci flow of the late epoch isotropic
and homogeneous spacetime. Phenomenologically speaking, the Ricci
flow and its counter term blurs the spacetime coordinate and equivalently
universally broadens the spectral lines (as the universe expanding
tracers). The broadening contributes a universal variance to the redshift,
thus the redshift-distance relation is modified at second order in
Taylor's series expanding the distance in powers of the redshift,
which gives rise to an equivalent accelerating expansion of the universe
as the quantum version Equivalence Principle asserts \citep{Luo2015Dark,Luo:2015pca,Luo:2019iby,Luo:2021zpi}.
Since the redshift variance (over the redshift mean squared) is independent
to the specific energies of the spectral lines, so they are seen universally
accelerating ``free-falling'' (in fact expanding), and the uniform
acceleration now is not merely a specific property of the spectral
lines, but measures and be interpreted as the universal property of
the quantum spacetime. 

When a distant earth observer measures the rotation velocities of
spiral galaxies at galactic long distance scale, the mechanism works
in a similar way. What the observer measures is not directly the rotation
velocity of the spiral galaxy but its Doppler (red and blue) shifts
induced broadening of the spectral lines (as the rotation tracers)
with respect to the ones in laboratory (as the starting reference).
As the galaxy is sufficiently redshifted at long distance scale, the
spectral lines themselves are intrinsically quantum broadened, interpreted
as the quantum variance or fluctuation of the distant spacetime coordinates
based on the quantum Equivalence Principle. Thus it gives additional
correction to the Doppler broadening and enhances the rotation velocities.

In treating the quantum fluctuation correction effect to spacetime
by the Ricci flow, it is useful to introduce an important and special
solution of the Ricci flow (or more general the Ricci-DeTurck flow
(\ref{eq:ricci-deturck}) for a density manifold) which only shrinks
the local size or volume of a manifold but its local shape unchanged,
named the Ricci Soliton (or more general Gradient Shrinking Ricci
Soliton of a density manifold) \citep{perelman2002entropy}. Its Ricci
curvature is proportional to the metric $R_{\mu\nu}=\frac{1}{2\tau}g_{\mu\nu}$
(or more general a gradient normalized Ricci curvature is proportional
to the metric)
\begin{equation}
R_{\mu\nu}-\nabla_{\mu}\nabla_{\nu}\log u=\frac{1}{2\tau}g_{\mu\nu}\label{eq:gsrs}
\end{equation}
where $\tau=t_{*}-t$ is a backwards Ricci flow parameter from a limit
scale $t_{*}$ (see section-II-A for details). The Gradient Shrinking
Ricci Soliton is a (temporary $t_{*}\neq0$ or final $t_{*}=0$) limit
spacetime configuration or a (local or global) fixed point in the
RG-sense, it (locally or globally) maximizes the Perelman's monotonic
functionals (at finite scale $t_{*}\neq0$ or IR limit scale $t=0$).
In some simple cases, including for examples, the homogeneous and
isotropic late epoch spacetime \citep{Luo:2019iby,Luo:2021zpi}, the
spatial inflationary early universe \citep{Luo:2021wdh}, and static
thermal equilibrium black hole \citep{Luo:2022statistics}, and the
topics concerned in the paper about the local galaxies, for all these
limit (or nearly limit) spacetime, the Gradient Shrinking Ricci Soliton
equation is more useful as simple examples. 

Further more, at the fundamental level, the proposed new framework
of quantum spacetime and related gravity seem to avoid several fundamental
difficulties that other approaches to quantum gravity typically face.
For examples, the renormalizability of the quantum gravity is the
renormalizability of the $d=4-\epsilon$ non-linear sigma model, or
at the Gaussian level correlates to the mathematical problem of the
convergence of the Ricci flow. The problem is solved based on the
works of Hamilton, Perelman and further developed by many other mathematicians,
especially after the discoveries of several monotonic functionals
for the Ricci flow in general dimensions and the generalization of
techniques to the non-compact and pseudo-Riemannian (Lorentzian) spacetime.
The Hilbert space of quantum gravity correlates to the classification
of spacetime geometries by using the Ricci flow approach, which is
fully solved in 3-space, and can be full understood in 4-spacetime
by using the Ricci flow approach without fundamental obstacle. The
unitarity of quantum gravity correlates to the problem of intrinsic
diffemorphism anomaly of the spacetime \citep{Luo:2022statistics},
which is given by the functional integral method for the quantum spacetime
reference frame and found deeply related to the thermodynamic nature
of the quantum spacetime. The local conformal stability of quantum
spacetime \citep{Luo:2022goc} correlates to the sign of the lowest
eigenvalue related to the F-functional of Perelman, and the collapsibility
of the quantum spacetime correlates to the finiteness of the W-functional
of Perelman. The background independence of quantum gravity correlates
to the initial (metric) condition independence of the Ricci flow,
i.e. the Ricci flow and quantum fluctuations about all general initial
background are on equal footings. The problem is considered fully
solved by the Perelman's formalism of the Ricci flow for general initial
condition of manifold (not be restricted on some special initial condition).
In the sense that the 2nd order quantum fluctuation of spacetime has
been important at low energy, it also has a completely different view
on the ``graviton'' w.r.t. the flat background. The ``graviton''
as the low energy excitation degrees of freedom of metric have been
averaged out in the Ricci flow and effectively contribute to the general
curved spacetime at certain scale, and hence it seems not to be a
good signature of missing-energy in particle collision, unlike some
high energy modified versions of quantum gravity.

As the quantum framework modifies the gravity at long distance scale
in such a very tight and rigid way, the primary objective of this
paper is to explore the weak and static gravity limit of the theory,
and to determine if MOND can be derived from the theory, and if so,
whether there is anything different or beyond MOND in the framework.

To avoid overlong pages, the general background of the quantum reference
frame and the Ricci flow is given in the introduction section, in
the next section, we skip the detail of the quantum reference frame
and direct starting from the partition function derived from it, which
can be found from the previous works \citep{Luo2014The,Luo2015Dark,Luo:2015pca,Luo:2019iby,Luo:2021wdh,Luo:2021zpi,Luo:2022goc,Luo:2022statistics}.
In the section II, we derived the low energy effective action of gravity
from the partition function. And in the subsequent sections, several
phenomenological consequences, e.g. the baryonic Tully-Fisher Relation
in section III, the radial acceleration relation in section IV and
``missing matter'' in section V are discussed. Finally we discuss
the relation between the theory and MOND and conclude the paper.

\section{Effective Gravity at Low Energy}

\subsection{Partition Function of Quantum Reference Frame and Pure Gravity}

By using the quantum frame fields described by a $d=4-\epsilon$ non-linear
sigma model, in \citep{Luo:2021zpi,Luo:2022goc} the author has derived
the partition function of a pure gravity in terms of the relative
Shannon entropy $\tilde{N}$ of the spacetime 4-manifold $M^{D=4}$
\begin{equation}
Z(M^{D})=e^{\lambda\tilde{N}(M^{D})-\frac{D}{2}-\nu}=\frac{e^{\lambda N(M^{D})-\frac{D}{2}-\nu}}{e^{\lambda N_{*}(M^{D})}}\label{eq:partition}
\end{equation}
where $D\equiv4$ and $\lambda=\frac{3H_{0}^{2}}{8\pi G}\approx\left(10^{-3}\mathrm{eV}\right)^{4}$
is the critical density of the universe. Up to a constant multiple,
it is in fact the inverse of Perelman's partition function \citep{perelman2002entropy}
(the inverse is not physical important) he used to deduce his thermodynamics
analogous functionals, although the underlying physical interpretation
and relation to gravity is unclear in his seminal paper. The partition
function is a proper starting point to pure gravity. Let us explain
some quantities appearing in the partition function as follows.

(1) $N$ and $N_{*}$ terms

The relative Shannon entropy $\tilde{N}(M^{4},t)$ is the Shannon
entropy $N(M^{4},t)$ w.r.t. the extreme value $N_{*}(M^{4},t_{*})$
given by a Gradient Shrinking Ricci Soliton (\ref{eq:gsrs}) as the
Ricci flow limit $t\rightarrow t_{*}$, i.e. $\tilde{N}=N-N_{*}$.
The quantum reference frame theory is described by a non-linear sigma
model in $d=4-\epsilon$, the trivialness of the homotopy group $\pi_{d}(M^{4})$
of the mapping of the non-linear sigma model makes the Ricci flow
globally singularity free, simply giving $t_{*}=0$, even though local
singularities may developed at finite $t_{*}\neq0$ during the Ricci
flow for some general initial spacetime. The Shannon entropy is given
by the $u$ density
\begin{equation}
N(M^{4},t)=-\int_{M^{4}}d^{4}Xu\log u,
\end{equation}
in which
\begin{equation}
u(X,t)=\frac{d^{4}x}{d^{4}X}\label{eq:u}
\end{equation}
is a volume ratio between a fiducial volume element $d^{4}x$ of the
lab's frame and the local spacetime (classical diffemorphism) invariant
volume element $d^{4}X\equiv dX^{0}dX^{1}dX^{2}dX^{3}\sqrt{|g|}$.
$u$ is a dimensionless manifold density at each point $X$ of the
Riemannian spacetime endowed with 2nd order moment quantum fluctuations,
satisfying the density normalization condition
\begin{equation}
\lambda\int d^{4}Xu=\lambda\int d^{4}x=1.\label{eq:u constraint}
\end{equation}
from which one can see that the full volume of the spacetime or the
fiducial lab's frame is about the inverse of the critical density
$O(1/\lambda)$, i.e. the volume of the universe, the characteristic
scale of the theory. Without loss of generality, one can attribute
the Ricci flow effects only to $\sqrt{|g|}$ and leaving $dX^{0}dX^{1}dX^{2}dX^{3}$
the fiducial volume, then $u$ is just the inverse of the Jacobian,
$u=1/\sqrt{|g|}$. By using the generalized Ricci-DeTurk flow (\ref{eq:ricci-deturck}),
and finally we obtain the conjugate-heat equation of the $u$ density
$\frac{\partial u}{\partial t}=\left(R-\varDelta\right)u$, where
$\varDelta$ is the Laplace-Beltrami operator of the spacetime, $R$
the scalar curvature of the spacetime. The equation is often written
in the form of a backwards flow along $d\tau=-dt$, i.e. 
\begin{equation}
\frac{\partial u}{\partial\tau}=\left(\varDelta-R\right)u,\label{eq:conjugate heat eq}
\end{equation}
so that the sign in front of $\varDelta$ is just correct to have
solution exist, as an analogous heat equation. The relative Shannon
entropy $N$ part in the exponential comes from the diffemorphism
anomaly when one performs functional integration over the frame fields
of the spacetime manifold \citep{Luo:2021zpi}, which also has profound
thermodynamic interpretation of spacetime \citep{Luo:2022statistics}. 

(2) $D/2$ term

The $D/2=2$ term arises from the classical action of a laboratory
(walls and clock) frame given by a non-linear sigma model, which is
considered classical, fiducial, rigid (volume-fixed), infinitely precise
and hence the quantum fluctuations of the laboratory frame are ignored.
We have
\begin{equation}
e^{-\frac{D}{2}}=\exp\left(-S_{cl}\right)=\exp\left(-\frac{1}{2}\lambda\int d^{4}xg^{\mu\nu}\partial_{a}x_{\mu}\partial_{a}x_{\nu}\right)=\exp\left(-\frac{1}{2}\lambda\int d^{4}xg^{\mu\nu}g_{\mu\nu}\right)
\end{equation}
where $x_{\mu}$ is the coordinates of the laboratory frame identified
with the classical value of the frame fields, i.e. $x_{\mu}=\langle X_{\mu}\rangle$
ignoring the quantum fluctuation. The quantum effects are all attributed
into the diffemorphism anomaly term $\lambda\tilde{N}$, thus the
effective action of the partition function can also be seen as an
anomaly induced action of the quantum non-linear sigma model.

(3) $\nu$ term

When the Wick rotated spacetime $M^{4}$ is topologically equivalent
to a 4-sphere $S^{4}$ which puts the space and time on an equal footing,
the Ricci flow will gradually deform the initial $M^{4}$ into an
isotropic and homogeneous $S^{4}$ at IR. In some cases, the flow
process is singularity-free, for example $M^{4}$ has already been
an isotropic and homogeneous spacetime with positive curvature (e.g.
late epoch universe), the Ricci flow just shrinks its volume but deforms
its shape. In other initial cases, the local irregularities of the
spacetime are not large and hence irrelevant, the flow then smooths
out them and flows the spacetime to the final IR isotropic and homogeneous
$S^{4}$ spacetime. In some general cases, singularities might developed
in some local places during the Ricci flow, then some local surgeries
are needed to remove the singularities and then continues the flow,
and finally it flows to several disconnected isotropic and homogeneous
$S^{4}$ as well, and an observer living in one of the disconnected
part of the universe, sees an isotropic and homogeneous $S^{4}$ universe.
Anyway, if one starts from the final $S^{4}$ with a proper choice
of a final density $u(t_{*})=u_{*}$ (up to a gauge), and traces it
backwardly by $\tau=t_{*}-t$, where $t_{*}$ is certain IR singular
scale of the flow (singularity-free case simply gives $t_{*}=0$)
and $t$ is the flow parameter of the forwards Ricci flow, from IR
$\tau=0$ backwardly to UV $\tau\rightarrow\infty$ ($S^{4}$ allows
the existence of the limit, i.e. $t\rightarrow-\infty$ called ancient
solution), we get
\begin{equation}
\nu(S^{4})\equiv\lambda\left[\tilde{N}(S^{4},\tau\rightarrow\infty)-\tilde{N}(S^{4},\tau=0)\right]=\log\tilde{V}(S^{4})\approx-0.8\label{eq:nu}
\end{equation}
which is calculated from the $\log$ of the Reduced Volume or Relative
Volume $\tilde{V}$ of Perelman \citep{perelman2002entropy,perelman2003ricci,perelman307245finite},
close to the observed $-\Omega_{\Lambda}$. $\nu$ is the finite difference
of $\lambda\tilde{N}(M^{4}\cong S^{4})$ between UV ($\tau\rightarrow\infty$)
and IR ($\tau=0$) following the analogous H-theorem of the Ricci
flow \citep{perelman2002entropy}. And because of the analogous irreversible
H-theorem (parallel to the a-theorem \citep{2011On} in 4d and c-theorem
\citep{c-theorem} in 2d in essential), the Ricci flow monotonically
maximized the Shannon entropy $N$ to $N_{*}$ at IR, so $\tilde{N}_{\tau=0}=0$
trivially. $\nu$ playing the role as a counter term and the fraction
of the cosmological constant, ensures the diffemorphism anomaly given
by $\lambda\tilde{N}(M^{4})$ can be completely canceled in the laboratory
frame up to UV scale, leaving only the classical action $D/2$ of
the laboratory frame, if the frame has been pre-assumed classical,
fiducial and rigid.

\subsection{Low Energy Expansion}

At low energy IR scale $t\rightarrow0^{-}$, or equivalently small
$\tau\rightarrow0^{+}$, the relative Shannon entropy can be expanded
in powers of $\tau$
\begin{equation}
\tilde{N}(M^{4},\tau)=\left.\frac{d\tilde{N}}{d\tau}\right|_{\tau\rightarrow0}\tau+O(\tau^{2})=\tau\left.\tilde{\mathcal{F}}\right|_{\tau\rightarrow0}+O(\tau^{2})
\end{equation}
in which $\tilde{N}=N-N_{*}$ and $\tilde{N}_{\tau=0}=0$. By using
the conjugate heat equation (\ref{eq:conjugate heat eq}) and Ricci
flow of the volume element $\frac{d}{d\tau}\left(d^{4}X\right)=R\left(d^{4}X\right)$,
we have
\begin{equation}
\frac{dN}{d\tau}=-\frac{d}{d\tau}\int_{M^{4}}d^{4}Xu\log u=\int_{M^{4}}d^{4}Xu\left(R+|\nabla\log u|^{2}\right)
\end{equation}
and the extreme value $N_{*}$ is given by a fundamental solution
of the Maxwell-Boltzmann type
\begin{equation}
\lim_{\tau\rightarrow0}u=u_{*}=\frac{1}{\lambda(4\pi\tau)^{2}}\exp\left(-\frac{1}{4\tau}|X-x|^{2}\right)
\end{equation}
and hence
\begin{equation}
\frac{dN_{*}}{d\tau}=-\frac{d}{d\tau}\int_{M^{4}}d^{4}Xu_{*}\log u_{*}=\frac{2}{\lambda\tau}
\end{equation}
so finally
\begin{equation}
\tilde{\mathcal{F}}=\int_{M^{4}}d^{4}Xu\left(R+|\nabla\log u|^{2}-\frac{2}{\tau}\right)
\end{equation}
is the normalized F-functional \citep{perelman2002entropy} of Perelman
$\tilde{\mathcal{F}}=\mathcal{F}-\mathcal{F}_{*}$ w.r.t. the maximized
value $\mathcal{F}_{*}=\frac{2}{\lambda\tau}$ in 4-spacetime.

At IR limit since $\lambda\int d^{4}Xu_{*}|\nabla\log u_{*}|^{2}=\frac{2}{\tau}$,
so the low energy effective action of the partition function (\ref{eq:partition})
can be written as
\begin{equation}
S_{eff}=-\log Z(M^{4})=\lambda\int_{M^{4}}d^{4}Xu_{*}\left[2-\tau R_{0}+\nu+O(R^{2}\tau^{2})\right],\quad(\tau\rightarrow0^{+}).\label{eq:EH+cc}
\end{equation}

From the Ricci flow equation (\ref{eq:ricci flow}) of the metric,
it is straightforward to give the flow equation of the scalar curvature
$\frac{\partial R}{\partial\tau}=-\varDelta R-2R_{\mu\nu}R^{\mu\nu}$.
Since in the IR flow limit $\tau\rightarrow0$, the Ricci curvature
is homogeneous and isotropic $\varDelta R_{0}=0$, and $R_{\mu\nu}(\tau\rightarrow0)=R_{0}g_{\mu\nu}$,
so the equation becomes $\frac{\partial R}{\partial\tau}=-\frac{2}{D}R^{2}=-\frac{1}{2}R^{2}$
in 4-spacetime, thus at small $\tau$ the solution is 
\begin{equation}
R_{\tau}=\frac{R_{0}}{1+\frac{1}{2}R_{0}\tau},\label{eq:scalar curvature flow}
\end{equation}
where $R_{0}=D(D-1)H_{0}^{2}=12H_{0}^{2}$. As a consequence the first
two terms in (\ref{eq:EH+cc}) can be interpreted as a scale($t$
or $\tau$)-dependent Einstein-Hilbert-like action 
\begin{equation}
2\lambda-\lambda R_{0}\tau=\frac{R_{\tau}}{16\pi G},
\end{equation}
in which $\lambda=\frac{3H_{0}^{2}}{8\pi G}=\frac{R_{0}}{32\pi G}$
has been used. Therefore the 3rd term $\lambda\nu$ as the counter
term of $\lambda\tilde{N}$ is naturally the cosmological constant
in the unit of the critical density $\lambda$, 
\begin{equation}
\lambda\nu=-\Omega_{\Lambda}\lambda=\frac{-2\Lambda}{16\pi G}.
\end{equation}
The 4th and subsequent terms $O(R^{n}\tau^{n})$ give high energy
corrections. In the IR limit, the fundamental solution of $u_{*}$
degenerates to a delta function, but considering that the spacetime
at IR limit is homogeneous and isotropic, and the general solution
of $u_{*}$ recovers a homogeneous density at IR limit, thus the integral
measure $d^{4}Xu_{*}$ recovers the classical invariant measure $d^{4}X\equiv dX^{0}dX^{1}dX^{2}dX^{3}\sqrt{|g|}$,
the density manifold $(M^{4},g,u)$ recovers the Riemannian manifold
$(M^{4},g)$. The introducing of matter to the quantum reference frame
and pure gravity can be found in \citep{Luo:2021zpi,Luo:2022goc}.
Finally, including the Lagrangian density of matters $\mathcal{L}_{M}$,
the effective action (\ref{eq:EH+cc}) can be rewritten as the standard
Einstein-Hilbert+Cosmological Constant (EH + CC) action,
\begin{equation}
S_{eff}=\int_{M^{4}}d^{4}X\left[\frac{R_{\tau}-2\Lambda}{16\pi G}+O(R^{2}\tau^{2})+\mathcal{L}_{M}\right],\quad(\tau\rightarrow0^{+})\label{eq:EH+CC+Mat}
\end{equation}
simply the scalar curvature and the matter Lagrangian becomes scale($t$
or $\tau$)-dependent.

Note that the standard EH + CC action has two input constants, the
Newton's constant $G$ and the cosmological constant (CC) $\Lambda$,
while strictly speaking, the action (\ref{eq:EH+cc}) has only one
input constant, the critical density $\lambda=\frac{3H_{0}^{2}}{8\pi G}$
as a combination of the two, which is a low energy density compared
with the Planck scale. The Planck scale instead is considered play
no fundamental role in the theory. It is clearly that in an ancient
solution configuration (limit $\tau\rightarrow\infty$ exists without
local singularities during the Ricci flow) the energy scale could
safely go beyond the Planck scale, the situation of the high energy
modification is beyond the scope of the paper when the whole partition
function (\ref{eq:partition}) is more appropriate to consider. At
the low energy limit (\ref{eq:EH+CC+Mat}), there are some crucial
observations worth stressing.

(i) The characteristic energy scale of the theory is as low as the
critical density and the cosmological constant, which gives a characteristic
scale or size to the universe in contrast to the naive predictions
of a quantum general relativity plus the cosmological constant (e.g.
leading to the cosmological constant problem). 

(ii) At low energy, $\tau=t_{*}-t\rightarrow0$, ($t_{*}=0$ is the
cosmic far infrared limit), there is a non-flat IR asymptotic spacetime
background $R_{0}=32\pi G\lambda$, even in the pure gravity theory
where matter is not included. The cosmological constant, as the counter
term of the Ricci flow, or equivalently, the constant asymptotic background
curvature, is the leading quantum correction at low energy. This not
only modifies the behavior of the asymptotic spacetime (i.e. the acceleration
of its expansion) but also affects the behavior of some long-distance
scale objects (i.e. the acceleration discrepancies in galaxies).

(iii) The next leading order quantum correction for the standard gravity
at low energy is that the observed scalar curvature $R_{\tau}$ and
metric in (\ref{eq:EH+CC+Mat}) are scale($t$ or $\tau$)-dependent.
Based on the Ricci flow of the scalar curvature $R$ and the metric
$g$, we have
\begin{equation}
R_{g}=\frac{R_{b}}{1-\frac{1}{2}R_{b}\Delta t}\quad\mathrm{and}\quad g_{\mu\nu}(t_{g})=g_{\mu\nu}(t_{b})\left(1-\frac{1}{2}R_{b}\Delta t\right),\quad\left(\Delta t=t_{g}-t_{b}>0\right)\label{eq:flow of R and g}
\end{equation}
where $-\infty<t_{b}<t_{g}\apprle0$. As a result, the gravity generated
by curvature $R_{b}$ and metric $g_{\mu\nu}(t_{b})$ is expected
to be distinct from the gravity generated by $R_{g}$ and $g_{\mu\nu}(t_{g})$.
The former is predicted from the baryonic matter at the local scale
$t_{b}$ ($b$ for baryon), which is a short-distance fiducial lab
scale used as a starting point for calibrating and extrapolating with
the optical law used for optical measurement of baryon. The latter
is at the galactic scale $t_{g}$ ($g$ for galaxy), which is related
to a relatively redshifted and quantum broadened long-distance scale
and observed by a distant observer.

(iv) If we assume the validity of the classical Einstein's equation,
then the flow change of the scalar curvature $\Delta R=R_{g}-R_{b}$
at different scales would give rise to the missing matter density
$\Delta\rho=\rho_{g}-\rho_{b}$ between these scales. In this sense,
the \textquotedblleft missing matter\textquotedblright{} is seen as
unavoidable in the theory when one predicts the effective masses from
the gravity coupled to them in different scales. However, since the
classical Einstein's equation is modified at long distances and is
no longer exact, the \textquotedblleft missing matter\textquotedblright{}
does not actually exist; they are merely an illusion created by the
Ricci flow of the curvature.

\subsection{Weak and Static Gravity Approximation}

In this subsection, first we derive a relativistic but weak fields
approximate action in the sense that the scalar curvature $R_{g}$
is low enough and comparable with the asymptotic background curvature
$R_{0}$ corresponding to the characteristic scale $\lambda$, which
describes how the galactic scale curvature $R_{g}$ is correlated
to the local scale baryonic matter Lagrangian density $\mathcal{L}_{M}$.
And then we study the static or non-relativistic approximation and
compare it with MOND. 

Dropping the higher order term and considering $\lambda=\frac{R_{0}}{32\pi G}$,
the effective action (\ref{eq:EH+CC+Mat}) can be rewritten in terms
of the dimensionless ratio $R_{\tau}/R_{0}=R_{g}/R_{0}$, in the unit
of $2\lambda$,
\begin{equation}
S_{eff}=\int_{M^{4}}d^{4}X\left[2\lambda\left(\frac{R_{g}}{R_{0}}-\frac{\Omega_{\Lambda}}{2}\right)+\mathcal{L}_{M}\right]=\int_{M^{4}}d^{4}X\left[2\lambda\frac{R_{g}}{R_{0}}\left(1-\frac{\Omega_{\Lambda}}{2}\frac{R_{0}}{R_{g}}\right)+\mathcal{L}_{M}\right].\label{eq:pre-approx}
\end{equation}
In the situation that gravity is strong, $R_{g}\gg R_{0}$, for example,
gravity well within the luminary region of galaxies or in the solar
system, the action can ignore the cosmological constant term $\frac{\Omega_{\Lambda}}{2}$
and recovers the well-tested standard Einstein's gravity, and hence
the static Newtonian gravity. 

However, at long distance scales, when the asymptotic gravity is weak,
i.e. the flow-averaged curvature $R_{g}$ is low and comparable to
the asymptotic background curvature $R_{0}$ at the coarse-graining
level, ratio $\frac{\Omega_{\Lambda}}{2}\frac{R_{0}}{R_{g}}\approx O(0.35)$
is still a small number less than 1, we could go from (\ref{eq:pre-approx})
to 
\begin{equation}
S_{eff}\approx\int_{M^{4}}d^{4}X\left[2\lambda\frac{R_{g}}{R_{0}}\frac{1}{\sqrt{1+\Omega_{\Lambda}\frac{R_{0}}{R_{g}}}}+\mathcal{L}_{M}\right]=\int_{M^{4}}d^{4}X\left[\frac{1}{16\pi G}\frac{R_{g}}{\sqrt{1+\Omega_{\Lambda}\frac{R_{0}}{R_{g}}}}+\mathcal{L}_{M}\right],\quad\left(R_{g}\approx O(R_{0})\right)\label{eq:approx-1}
\end{equation}
under the approximation condition
\begin{equation}
R_{\tau}=R_{g}\approx O(R_{0}).\label{eq:coarse-graining approximation}
\end{equation}

The purpose of the approximation is to incorporate the global effects
of the cosmological constant or background curvature into a local
gravitational system, which is more suitable as a starting action
to study the modified law of local gravity. In the action, the matter
term is still the dominant factor in the effective curvature, as if
the cosmological constant were absent in the usual consideration of
local gravitational binding systems. However, the effect of the cosmological
constant has also become increasingly important at the outskirts of
the binding systems, where the curvature becomes low, as indicated
by its absorption into the effective scalar curvature under (\ref{eq:coarse-graining approximation}).
In this approximation, the contribution of the cosmological constant
just effectively gives an extra factor $\left(1+\Omega_{\Lambda}R_{0}/R_{g}\right)^{-1/2}$
to the scalar curvature. 

Indeed, there are many other functions that can be used to approximate
the action (\ref{eq:pre-approx}). The form of the approximation (\ref{eq:approx-1})
we choose is to take MOND as a phenomenological reference. In fact,
there are also other allowed forms of interpolation functions in MOND
that work about equally well phenomenologically. However, the accuracy
of current cosmic measurements is not good enough to uniquely fit
or to distinguish them very well phenomenologically. Nevertheless,
if the theory could (at leading order) give one of the allowed interpolating
functions of MOND, then we could say that the theory could qualitatively
give a similar transition trend of MOND at least at the leading order
within the allowed range of current observational accuracy. The form
of the approximation (\ref{eq:approx-1}) is the most direct and closest
form in all possible and allowed interpolating functions of MOND appearing
in literature that can be derived from the action (\ref{eq:pre-approx})
at the leading order. If one chooses other forms and plots them by
curves \textquotedbl at leading order\textquotedbl , they will work
about equally well to fit data.

At this point, we further consider the weak $\Phi\ll1$ and static
$\dot{\Phi}\approx0$ gravity approximation
\begin{equation}
g^{00}\approx-\left(1-2\Phi\right),\quad g^{ij}\approx(1+2\Phi)\delta_{ij}.
\end{equation}
where $\Phi$ without subscript is the Newtonian potential in weak
and static approximation from the metric in general scale. The potential
taking a subscript e.g. $\Phi_{g}$ is the observed potential at the
specific galactic scale, $\Phi_{b}$ at the baryonic scale. Roman
letters $i,j$ are used for spatial indices. In the limit, scalar
curvature $R$ includes $\varDelta\Phi$, and second order ones such
as $(\nabla\Phi)^{2}$ and $\Phi\varDelta\Phi$. To derive the fields
equation, at the linear level, the terms $\Delta\Phi$ becomes immaterial
in the action, as a complete derivative, and hence we are left with
terms $(\nabla\Phi)^{2}$ (note that $\Phi\varDelta\Phi$ terms is
also $(\nabla\Phi)^{2}$ up to a derivative). At linear level $R_{g}$
can be replaced by $2|\nabla\Phi_{g}|^{2}$ without significantly
change the classical fields equation of Newtonian potential we concern
in the weak and static gravity limit.

In contrast to the fact that the gravity part of the Lagrangian as
geometric quantity varies with the Ricci flow, the matter part $\mathcal{L}_{M}$
is assumed not, so we write down local $\mathcal{L}_{M}$ at the conventional
fixed baryonic scale. The proper stress tensor, which does not contain
metric and using the covariant index, is given by
\begin{equation}
T_{00}=\rho_{M},\quad T_{ij}\approx0,
\end{equation}
in which we have considered the proper velocity and pressure of the
baryonic matters in galaxy are low. The proper stress tensor is coupled
to the local metric $g^{\mu\nu}(t_{b})$ or gravity potential $\Phi_{b}$
at the baryonic scale. Thus it is the conventional baryonic scale
$\mathcal{L}_{M}$ inserting into the total effective action
\begin{equation}
\mathcal{L}_{M}\approx\frac{1}{2}g_{b}^{00}T_{00}=-\frac{1}{2}\left(1-2\Phi_{b}\right)\rho_{M}=-\frac{1}{2}\left(1-2\Phi_{g}\right)\left(1-\frac{1}{2}R_{b}\Delta t\right)\rho_{M}
\end{equation}
in which the local comoving metric $g_{b}^{00}$ has been transformed
to the galactic scale metric $g_{g}^{00}$ observed by a distant observer
via the Ricci flow 
\begin{equation}
g^{00}(t_{g})=\frac{g^{00}(t_{b})}{1-\frac{1}{2}R_{b}\Delta t}.
\end{equation}
Finally we have the action for the potential observed at the galaxy
scale coming from a local baryonic density
\begin{equation}
S_{eff}=\int d^{4}X\left[\frac{|\nabla\Phi_{g}|^{2}}{8\pi G}\frac{1}{\sqrt{1+\Omega_{\Lambda}\frac{6H_{0}^{2}}{|\nabla\Phi_{g}|^{2}}}}-\frac{1}{2}\left(1-2\Phi_{g}\right)\left(1-\frac{1}{2}R_{b}\Delta t\right)\rho_{M}\right],\quad\left(|\nabla\Phi_{g}|\approx O(H_{0})\right).\label{eq:weak static action}
\end{equation}

Compared with the action of the standard Newtonian gravity (Poisson
action), there are two extra factors appear. First is the factor $\left(1+\Omega_{\Lambda}\frac{6H_{0}^{2}}{|\nabla\Phi_{g}|^{2}}\right)^{-1/2}$
modifying the quadratic term $|\nabla\Phi_{g}|^{2}$, which will give
a transition trend similar with MOND. The second is the factor $1-\frac{1}{2}R_{b}\Delta t$
modifying the Newtonian potential coupled to the baryonic density
source, which will play the role of a scale dependent ``missing matter''.
In the Newtonian limit when these two extra factors are trivially
unity, the effective action can recover back to obtain the Poisson
equation of Newtonian potential $\Phi_{N}\approx\Phi_{g}\approx\Phi_{b}$
\begin{equation}
\varDelta\Phi_{N}=4\pi G\rho_{M}
\end{equation}
when the scale difference is small $\Delta t\rightarrow0$ and the
gravity is much stronger than the background one $|\nabla\Phi_{N}|^{2}\gg H_{0}^{2}$.

Further, there is a transition trend away from Newtonian gravity in
the low curvature region $R_{g}\approx O(R_{0})$, where the gravity
appears to be strongly modified as the deep-MOND limit. More precisely,
the limit requires $R_{g}\ll R_{0}$ which is beyond the approximation
(\ref{eq:coarse-graining approximation}) and hence unreachable, if
we extrapolate the approximation region to $R_{g}\ll R_{0}$, i.e.
\begin{equation}
\varDelta\Phi_{g}\ll|\nabla\Phi_{g}|^{2}\ll6\Omega_{\Lambda}H_{0}^{2}=2\Lambda\label{eq:strong modified condition}
\end{equation}
but we have an extrapolating deep-MOND action from (\ref{eq:weak static action})
\begin{align}
S_{eff} & \rightarrow\int d^{4}X\left[\frac{1}{8\pi G\sqrt{2\Lambda}}|\nabla\Phi_{g}|^{3}-\frac{1}{2}\left(1-2\Phi_{g}\right)\left(1-\frac{1}{2}R_{b}\Delta t\right)\rho_{M}\right],\quad(|\nabla\Phi_{g}|^{2}\ll2\Lambda).\label{eq:strong modified action}
\end{align}
Since little matter locate far from the galaxy center, so in (\ref{eq:strong modified condition}),
$|\nabla\Phi_{g}|^{2}\gg\varDelta\Phi_{g}\approx0$ is used and $|\nabla\Phi_{g}|^{2}$
dominating the curvature $R_{g}\approx2|\nabla\Phi_{g}|^{2}+\varDelta\Phi_{g}$.

However, it is worth stressing that although $a_{N}\sim|\nabla\Phi_{N}|$
could be very small, the condition of the deep-MOND action (\ref{eq:strong modified action})
$|\nabla\Phi_{g}|^{2}\sim a^{2}\ll2\Lambda$ is unreachable at the
coarse-graining level, since in the limit action, the approximation
condition (\ref{eq:coarse-graining approximation}) and action (\ref{eq:approx-1})
fail globally. The action (\ref{eq:weak static action}) shows a transition
trend at low curvature region $R_{g}\approx O(R_{0})$ towards the
extrapolating limit (\ref{eq:strong modified action}). 

To investigate the region when the baryonic matter density $\rho_{M}$
is much lower than the critical density $\lambda$, or equivalently
the limit $a_{N}\rightarrow0$. Comparing the Poisson action and the
standard EH + CC action (\ref{eq:EH+CC+Mat}), if we effectively consider
$\frac{1}{2}\left(R_{g}-2\Lambda\right)\sim|\nabla\Phi_{N}|^{2}\sim a_{N}^{2}$,
then as $a_{N}\rightarrow0$ it leads to a universal acceleration
lower bound $R_{g}\rightarrow2\Lambda\sim O(R_{0})$. If the observed
acceleration is related to the curvature $a^{2}\sim|\nabla\Phi_{g}|^{2}=\frac{1}{2}R_{g}$,
it leads to a minimal observed acceleration $a\rightarrow a_{min}\sim\sqrt{\Lambda}$
corresponding to the universal cosmic expanding acceleration. Here
``$\sim$'' means up to certain correction factor. The existence
of the minimal acceleration might occur when crossing over from a
local galaxy binding gravitational system to the cosmological expanding
system. It is natural to imagine, since the rotational acceleration
of a galaxy is measured by the Doppler broadening of the (e.g. 21cm)
spectral lines at the outskirt of the galaxy, so when the acceleration
is low enough, the line broadening will be finally dominated by the
universal quantum broadening due to the Ricci flow of spacetime so
that the acceleration seems to achieve a universal value related to
the acceleration expansion of the spacetime. In other equivalent words,
when the binding gravity is so low at the outskirt of the galaxy that
the satellites of the galaxy will escape from its binding and experience
the universal cosmic expanding acceleration given by the cosmological
constant.

The possible existence of a minimal acceleration differs from MOND
at the deep-MOND limit when $\rho_{M}\ll\lambda$ or $a_{N}\ll\sqrt{\Lambda}$,
which predicts that the rotation acceleration of galaxy can really
reach a low value in the deep-MOND limit since the cosmological constant
and the background acceleration expansion of the spacetime are in
fact not taken into account in MOND. There are certain observational
hints \citep{Federico2017One} for such a coarse-grained ``acceleration
floor'' (see Figure 1 citing from \citep{Federico2017One}), especially
for those ultrafaint dSphs having some tension with the prediction
of MOND. The acceleration floor is fitted to be about $(9.2\pm0.2)\times10^{-12}m/s^{2}$
which is lower than the order of the cosmological constant $O(\Lambda)$.
There are also possible explanations to weaken such tension, for instance,
the transforming from the cosmological constant $\sqrt{\Lambda}$
to an acceleration $a$ is not simply equal. Some nontrivial extra
factors due to the Ricci flow might contribute, a reasonable renormalization
factor is $1-\frac{1}{2}R_{b}\Delta t$, in which $1-\frac{1}{2}R_{b}\Delta t\approx(0.13\sim0.18)$
(see later) is for the renormalization of the coupling between matter
and potential/acceleration, so the minimal acceleration might be suppressed
to $a_{min}\sim\left(1-\frac{1}{2}R_{b}\Delta t\right)\sqrt{\Lambda}\sim a_{0}\sim O(10^{-10}m/s^{2})$.
To fit the observed order $O(10^{-11}m/s^{2})$, further correction
is needed, for example, the renormalization of curvature. Actually
if the acceleration is able to reach the deep-MOND limit, then as
$a\rightarrow0$, the final constant rotation curve $v^{2}=a\cdot r$
of a galaxy would be extended to infinity $r\rightarrow\infty$, which
is of course impossible from the cosmological viewpoint. There would
be a maximal limit radius of rotating satellites beyond which the
curvature and related acceleration achieve minimum, rather than reaches
the deep-MOND limit $a\rightarrow0$. In general, since in the framework
the acceleration (defined by the 2nd spacetime derivative of the coordinate)
of a test satellite at the outskirt of a galaxy is affected by the
variance or 2nd moment fluctuation of the spacetime coordinate, if
the Ricci flow and the 2nd moment quantum fluctuation of spacetime
coordinate at long distance scale are intrinsically unavoidable, and
the rotation velocity and related acceleration of the satellites are
essentially measured by the broadening (variance) of the spectral
lines, there would very possibly exist a final broadening limit for
distant spectral lines and a fundamental lower bound for the acceleration
of the test satellites.

In fact the most salient aspect of MOND might not be the explicit
Lagrangian but the scaling relation towards the deep-MOND limit, so
in this sense the transition trend of (\ref{eq:weak static action})
without the deep-MOND limit has been sufficient to reproduce the core
results of MOND, whose observed data points have rotating acceleration
be larger than about $10^{-11}m/s^{2}$ (shown by the blue square
data points in Figure 1). It is not necessary to actually reach the
deep-MOND limit, as is shown by the blue points in the Figure 1 that
they are just within the range of $(10^{-11}\sim10^{-9})m/s^{2}$
of equivalent order of $O(\sqrt{R_{0}})\sim O(\sqrt{\Lambda})$ or
normalized acceleration $O(a_{0})$. In fact through measuring the
rotation curve of a galaxy by conventional optical method (e.g. 21cm
spectral lines), no rotating acceleration data points are really low
enough to reach the deep-MOND limit $a\ll10^{-11}m/s^{2}$, otherwise
it is also natural to imagine that the satellites of galaxy will escape
from its local gravitational binding and to experience the universal
cosmic expanding acceleration.

\subsection{The Interpolating Lagrangian Function}

The transition trend crossing over from the Newtonian Limit to the
unreachable asymptotic deep-MOND limit is given by the interpolating
Lagrangian function, which will gives rise to the blue data points
of finite range in Figure 1. The gravity part Lagrangian in (\ref{eq:weak static action})
is simplified as
\begin{equation}
\mathcal{L}_{gra}=\frac{2\Lambda}{8\pi G}\frac{|\nabla\Phi_{g}|^{2}}{2\Lambda}\frac{1}{\sqrt{1+\frac{2\Lambda}{|\nabla\Phi_{g}|^{2}}}}=\frac{\Lambda}{4\pi G}F(x),\quad\left(x=\frac{|\nabla\Phi_{g}|^{2}}{2\Lambda}\right),\label{eq:interpolating function by Lambda}
\end{equation}
where $\Lambda=\Omega_{\Lambda}\frac{3H_{0}^{2}}{8\pi G}$ is the
cosmological constant and 
\begin{equation}
F(x)=\frac{x}{\sqrt{1+x^{-1}}}=\begin{cases}
x & x\gg1,\quad(\textrm{Newtonian\;Limit})\\
x^{3/2} & x\ll1,\quad(\textrm{Unreachable\;Asymptotic\;deep-MOND\;Limit})
\end{cases}
\end{equation}
 is a interpolating Lagrangian function. 

To see its relation to the MOND, the curvature ratio $R/R_{0}$ here
is replaced by the acceleration ratio $a/a_{0}$ in MOND, if we take
$a=|\nabla\Phi_{g}|$ and 
\begin{equation}
a_{0}=\frac{2}{3}\sqrt{2\Lambda},\label{eq:a0-1}
\end{equation}
which has a correct order but several times larger than the observed
value. The interpolating Lagrangian can be rewritten as a interpolating
Lagrangian similar with a non-relativistic generalization of MOND
\citep{Bekenstein:1984tv}
\begin{equation}
\mathcal{L}_{gra}=\frac{a_{0}^{2}}{4\pi G}\frac{|\nabla\Phi_{g}|^{2}}{a_{0}^{2}}\frac{1}{\sqrt{1+\frac{9}{4}\frac{a_{0}^{2}}{|\nabla\Phi_{g}|^{2}}}}=\frac{a_{0}^{2}}{4\pi G}f\left(\frac{|\nabla\Phi_{g}|^{2}}{a_{0}^{2}}\right)\label{eq:interpolating MOND}
\end{equation}
where
\begin{equation}
f(y)=\frac{y}{\sqrt{1+\frac{9}{4}y^{-1}}}=\begin{cases}
y & y\gg1,\quad(\mathrm{Newtonian\;Limit})\\
\frac{2}{3}y^{3/2} & y\ll1,\quad(\mathrm{Unreachable\;Asymptotic\;deep-MOND\;Limit})
\end{cases},\quad\left(y=\frac{a^{2}}{a_{0}^{2}}\right).
\end{equation}
So at leading order, the theory predicts a critical acceleration $a_{0}$
having a correct order $O(\sqrt{\Lambda})$. The next leading order
corrections to $a_{0}$ is considered in the baryonic Tully-Fisher
relation in the next section.

\section{The Baryonic Tully-Fisher Relation}

The baryonic Tully-Fisher relation $v^{4}=GMa_{0}$ is an empirical
tight relationship between the baryonic mass $M$ (including the visible
and invisible baryonic mass in gas) and its asymptotic flat rotation
velocity $v$ of disk galaxies, in which $G$ is the Newton's constant
and $a_{0}\approx1.2\times10^{-10}m/s^{2}\approx\frac{\sqrt{\Lambda}}{\left(6\sim8\right)}$
is a nearly universal constant interpreted as a critical acceleration
in MOND. The baryonic Tully-Fisher relation, especially the 4th power
of the rotation velocity can not be naturally explained in a cold
dark matter model, in which a naive expectation is about 3rd power. 

Similar with MOND, as long as the deep-MOND limit can be asymptotic
approached (not necessarily be reached), the Tully-Fisher relation
can be produced from the MOND-like transition trend (\ref{eq:weak static action})
and its asymptotically approached limit in a natural way. Based on
the transition trend to the asymptotic deep-MOND Limit (\ref{eq:strong modified action}),
the derivation of the relation is basically similar with MOND with
a little correction from the Ricci flow.

A simple way to derive the baryonic Tully-Fisher relation is to take
the analogy of deriving a non-relativistic dispersion relation from
a relativistic one. Comparing the Poisson action and the Einstein-Hilbert
+ CC action, we can see that the effective baryonic (Newtonian) acceleration
is $\frac{1}{2}\left(R_{g}-2\Lambda\right)=|\nabla\Phi_{N}|^{2}\sim a_{N}^{2}$,
and if the observed acceleration is simply taken as $a^{2}\sim|\nabla\Phi_{g}|^{2}=\frac{1}{2}R_{g}^{2}$,
then
\begin{equation}
a_{N}\sim\sqrt{\frac{1}{2}R_{g}-\Lambda}\sim\sqrt{a^{2}-\Lambda}
\end{equation}
The relation is in analogy with the relativistic dispersion relation
$E=\sqrt{p^{2}+m^{2}}$, where we have the analogies $\Lambda\sim-m^{2}\sim a_{0}^{2}$,
the observed acceleration $a^{2}\sim p^{2}$, and baryonic Newtonian
acceleration $a_{N}\sim E$. In a non-relativistic, $p^{2}\lesssim m^{2}$,
we obtain a non-relativistic trend $E\sim\frac{p^{2}}{2m}$ up to
a rest mass constant. The low matter density limit $\rho_{M}\ll\lambda$
is in analogy with the non-relativistic limit, (but note the minus
sign in front of $\Lambda$, so $a^{2}$ could not actually lower
than $\Lambda$, otherwise $a^{2}-\Lambda$ in the squared root would
become negative), but when $a^{2}\rightarrow O(\Lambda)$ or $R_{g}\approx O(R_{0})$,
the transition trend of the magnitude of the acceleration is given
by the MOND dynamics
\begin{equation}
\left|a_{N}\right|\sim\frac{a^{2}}{2\sqrt{\Lambda}},\quad\left(a^{2}\approx O(\Lambda)\right)
\end{equation}
up to a background acceleration constant. If one considers the limit
rotational velocity $v$ of a galaxy at radius $r$, i.e. $a=\frac{v^{2}}{r}$,
and Newtonian acceleration at the same radius, $a_{N}=\frac{GM}{r^{2}}$,
then the transition trend $|a_{N}|\sim\frac{a^{2}}{2\sqrt{\Lambda}}$
gives rise to the baryonic Tully-Fisher relation $v^{4}\sim2GM\sqrt{\Lambda}\sim GMa_{0}$
although the deep-MOND limit $a\ll O(\Lambda)$ is unreachable. 

If the deep-MOND region is really unreachable as $\rho_{M}\rightarrow0$
or $a_{N}\rightarrow0$, there is a maximal radius $r_{max}$ corresponding
to the minimal acceleration floor $a_{min}\sim O(a_{0})\sim O(\sqrt{\Lambda})$,
i.e. $\frac{GM}{r_{max}^{2}}\sim a_{min}$, when $r\gtrsim r_{max}$
the outskirt satellites finally escape from the binding of the local
galaxy and experience the background acceleration expansion. Given
the rotating platform velocity $v\left(r\lesssim r_{max}\right)$
of the satellites, relation $v^{2}=a_{min}\cdot r_{max}$ also gives
the baryonic Tully-Fisher relation $v^{2}\sim\sqrt{GMa_{min}}\sim\sqrt{GMa_{0}}$. 

The above hand waving derivations suggest a simple picture that around
(rather than deep below) $O(\sqrt{\Lambda})$, the asymptotic behavior
of acceleration discrepancy in MOND and the Tully-Fisher relation
might qualitatively come from the local effect of the cosmological
constant. 

To find the precise value of $a_{0}$, we could derive the baryonic
Tully-Fisher relation in another more precise approach, in which the
normalization of the coupling between matter and the potential can
be taken into account. We consider the Euler-Lagrangian equation $\nabla_{i}\frac{\delta\mathcal{L}}{\delta\nabla_{i}\Phi_{g}}-\frac{\delta\mathcal{L}}{\delta\Phi_{g}}=0$
of the limit Lagrangian (\ref{eq:strong modified action})
\begin{equation}
\frac{3}{8\pi G\sqrt{2\Lambda}}\vec{\nabla}\cdot\left(|\vec{\nabla}\Phi_{g}|\vec{\nabla}\Phi_{g}\right)-\left(1-\frac{1}{2}R_{b}\Delta t\right)\rho_{M}=0.
\end{equation}
By using the Poisson equation $\varDelta\Phi_{N}=4\pi G\rho_{M}$,
the baryonic matter density can be replaced by the Newtonian potential
\begin{equation}
\frac{3}{2\sqrt{2\Lambda}\left(1-\frac{1}{2}R_{b}\Delta t\right)}\vec{\nabla}\cdot\left(|\vec{\nabla}\Phi_{g}|\vec{\nabla}\Phi_{g}\right)=\vec{\nabla}\cdot\vec{\nabla}\Phi_{N}.
\end{equation}
Naturally to assume that the gravity potential $\Phi_{g}$ and $\Phi_{N}$
are spherical symmetric and curl-less, so 
\begin{equation}
\frac{3}{2\sqrt{2\Lambda}\left(1-\frac{1}{2}R_{b}\Delta t\right)}|\vec{\nabla}\Phi_{g}|\vec{\nabla}\Phi_{g}=\vec{\nabla}\Phi_{N}=|\nabla_{r}\Phi_{N}|\vec{e}_{r},
\end{equation}
where $\vec{e}_{r}$ is a unit vector in the radial direction. Therefore
\begin{equation}
\frac{3}{2\sqrt{2\Lambda}\left(1-\frac{1}{2}R_{b}\Delta t\right)}|\nabla_{r}\Phi_{g}|^{2}=|\nabla_{r}\Phi_{N}|=\frac{GM}{r^{2}},
\end{equation}
where $M$ is the total mass of the baryonic matters, simply assuming
that baryonic matters are centrosymmetric distributed (it is true
in most galaxies with a little distribution correction). Considering
e.g. spiral galaxies, a star at the outskirt of the galaxy, rotating
with an observed rotating platform velocity $v$, radius $r$ and
rotation radial acceleration $a_{r}=\frac{v^{2}}{r}=-\nabla_{r}\Phi_{g}$,
so we have
\begin{equation}
a_{r}^{2}=\left(\frac{v^{2}}{r}\right)^{2}=\frac{2}{3}\sqrt{2\Lambda}\left(1-\frac{1}{2}R_{b}\Delta t\right)\frac{GM}{r^{2}}
\end{equation}
in which the correction factor $1-\frac{1}{2}R_{b}\Delta t$ can be
considered as a renormalization to the coupling between matter and
potential or acceleration $a_{N}=\frac{GM}{r^{2}}$. So
\begin{equation}
v^{4}=\frac{2}{3}\sqrt{2\Lambda}\left(1-\frac{1}{2}R_{b}\Delta t\right)GM,
\end{equation}
which is just the baryonic Tully-Fisher relation with a critical acceleration
\begin{equation}
a_{0}=\frac{2}{3}\sqrt{2\Lambda}\left(1-\frac{1}{2}R_{b}\Delta t\right),\label{eq:scale dependent a0}
\end{equation}
The predicted critical acceleration is the universal part (\ref{eq:a0-1})
plus a scale dependent correction term. Although there are many possible
corrections to the universal part (\ref{eq:a0-1}), such as the mass
distribution (not point mass) correction in galaxies, the scale dependent
part $R_{b}\Delta t$ is also a possible correction rigidly predicted
in the Ricci flow of spacetime which may be important amount other
contributions. To fit the observed value $a_{0}\approx\frac{\sqrt{\Lambda}}{\left(6\sim8\right)}$
for spiral galaxies at the galactic scale $t_{g}$ w.r.t. $t_{b}$,
if we attempt to attribute all the discrepancy to the scale correction,
it equivalent to require a rough value
\begin{equation}
\frac{1}{2}R_{b}\Delta t=\frac{1}{2}R_{b}(t_{g}-t_{b})\approx\left(0.82\sim0.87\right).\label{eq:correction of a0}
\end{equation}

As is shown in eq.(\ref{eq:flow of R and g}), the correction factor
$1-\frac{1}{2}R_{b}\Delta t\approx(0.13\sim0.18)$ also renormalizes
the scalar curvature $R_{g}$, in this case, the action (\ref{eq:pre-approx})
differs from the standard EH action in $R_{b}$ by about $\frac{R_{g}-R_{b}}{R_{b}}\approx\left(4.6\sim6.7\right)$
times. Note that the approximation in the action (\ref{eq:approx-1})
deviates from (\ref{eq:pre-approx}) by about 10\%. In this sense,
the approximated action (\ref{eq:approx-1}) is closer to MOND than
the standard Einstein-Hilbert theory.

The correction term is seen scalar curvature $R_{b}$ and scale difference
$\Delta t$ dependent, but here the combination of the two roughly
takes a constant value and varies little with $\Delta t$. There indeed
exists a special reason for the case. Since the Ricci flow is highly
non-linear, the spacetime under the Ricci flow not always be smoothed
out as the linear heat equation. During the flow, some spacetime regions
with high curvature might develop local singularities at certain finite
singular scale $t_{*}\neq0$ (although the global singularity could
avoid by the topological consideration in the dimension of the base
space of the quantum reference frame). Galaxies as relative high curvature
local regions in the universe might play such role, meaning that $t_{g}$
might be close to a finite singular scale $t_{g}\approx t_{*}$. Near
the singular scale, the galaxy region roughly resembles a local Shrinking
Ricci soliton-like spacetime configuration (\ref{eq:gsrs}), for which
the combination of $R$ and $\tau=t_{*}-t_{b}$ is just about a constant
of order one \citep{perelman2002entropy}, and so $R_{b}\Delta t\sim O(1)$.
As the flow limit configuration, the Shrinking Ricci soliton flow
equation (\ref{eq:gsrs}) only deforms its local volume but local
shape so it is a self-similar configuration, and it is for this reason
the combination $R_{b}\Delta t$ could weakly depend on its scale.
According to the explanation, the nearly constant value of the correction
term $R_{b}\Delta t$ might imply that galaxy (and surrounding spacetime
configuration) is close to and resembles a Shrinking Ricci Soliton
configuration, and hence $a_{0}$ is roughly universal for different
galaxies. In fact whether $a_{0}$ is universal for all galaxies is
still in controversy \citep{Rodrigues:2018duc,Li:2018tdo}. To our
knowledge, some obscure evidences show $a_{0}$ might correlate with
the central surface brightness \citep{10.1093/mnras/stu100}, if so,
it might be roughly understood as a possible correlation to the curvature,
while there is still little evidence \citep{2012The} for a redshift
evolution of $a_{0}$. 

\section{The Radial Acceleration Discrepancy of Galaxies}

The baryonic Tully-Fisher relation is a simple scaling relation with
no apparent dependence on other properties like size or surface brightness
of galaxies, and it mainly tests the asymptotic low curvature or low
acceleration limit at limit galaxian radius. To detail test the interpolating
Lagrangian function for other intermediate accelerations at different
radii in different galaxies, a sharp empirical relation, the Radial
Acceleration Relation of galaxies, by McGaugh, Lelli and Schombert
is worth considering. The relation subsumes and generalizes the Tully-Fisher
relation. The radial acceleration $a$ are measured followed by 2693
points in 153 galaxies with very different morphologies, mass, sizes
and gas fractions, distributed up to the range $a_{obs}\gtrsim O(10^{-11}m/s^{2})$
shown in the Figure 1. It is surprisingly found their strong correlations
with that expected from the only baryonic acceleration $a_{bar}$
without any particular halo distribution model of the dark matter.
The blue data points (ignoring the ``acceleration floor'') of Figure
1 is fitted \citep{Stacy2016Radial,Federico2017One} as 
\begin{equation}
a=\frac{a_{N}}{1-e^{-\sqrt{\frac{a_{N}}{a_{0}}}}}\label{eq:radial acceleration fit}
\end{equation}
where $a_{0}\approx1.2\times10^{-10}m/s^{2}\approx\frac{\sqrt{\Lambda}}{\left(6\sim8\right)}$
is again empirically the critical acceleration. The circles and diamonds
in the Figure 1 given by the dwarf spheroidals (dSphs) (distinguish
between MW and M31 satellites) data points shows a possible ``acceleration
floor'' at about $(9.2\pm0.2)\times10^{-12}m/s^{2}$. 

We plot the radial acceleration discrepancy by the black curve in
the Figure 2. Note in the eq.(\ref{eq:weak static action}) that the
coupling between matter and potential (and acceleration) is normalized
by a factor $1-\frac{1}{2}R_{b}\Delta t\approx(0.13\sim0.18)$. Let
both $a$ and $a_{N}$ coming from potential $\Phi_{g}$ and $\Phi_{N}$
are normalized by such a factor, then the observed acceleration is
\begin{equation}
a=\left(1-\frac{1}{2}R_{b}\Delta t\right)|\nabla\Phi_{g}|=\left(1-\frac{1}{2}R_{b}\Delta t\right)\sqrt{\frac{1}{2}R_{g}}
\end{equation}
 and the effective Newtonian acceleration
\begin{equation}
a_{N}=\left(1-\frac{1}{2}R_{b}\Delta t\right)|\nabla\Phi_{N}|=\left(1-\frac{1}{2}R_{b}\Delta t\right)\sqrt{\frac{1}{2}\left(R_{g}-2\Lambda\right)}=\sqrt{a^{2}-\left(1-\frac{1}{2}R_{b}\Delta t\right)^{2}\Lambda}=\sqrt{a^{2}-\frac{9}{8}a_{0}^{2}},
\end{equation}
in which by $a_{0}=1.2\times10^{-10}m/s^{2}$ is used. So the observed
acceleration $a=\sqrt{a_{N}^{2}+\frac{9}{8}a_{0}^{2}}\approx\sqrt{a_{N}^{2}+a_{0}^{2}}$
is plotted by the black curve in the Figure 2, in which the curve
is away from the Newtonian prediction (orange dotted line) at about
$O(\sqrt{R_{0}})$, and has an asymptotic acceleration floor at about
$a_{0}\sim10^{-10}m/s^{2}$ (dash-dot line). $a_{0}$ can be seen
as the normalization version corresponding to the background curvature
$\sqrt{R_{0}}$ (dot-dash line), so naively there should be no accelerations
below the acceleration floor, i.e. no curvatures below the asymptotic
background curvature $R_{0}$ by the coarse-graining effect of the
Ricci flow to the spacetime. However, there are indeed accelerations
data points below $a_{0}$, and the acceleration floor $a_{min}$
in Figure 1 is seen lower than $a_{0}$. A possible reason might be
that further normalization corrections are also needed, for instance,
the renormalization to the curvature sqaure-root $a\sim\sqrt{R}$
itself also by the factor $\left(1-\frac{1}{2}R_{b}\Delta t\right)^{1/2}$,
(see (\ref{eq:flow of R and g})), to about $a_{min}^{re}\sim\left(1-\frac{1}{2}R_{b}\Delta t\right)^{1/2}a_{0}\sim10^{-11}m/s^{2}$,
which is about the order of the observed acceleration floor in Figure
1. But unfortunately if we further renormalize and lower the minimal
acceleration, the transition trend would further deviate from the
fitting (\ref{eq:radial acceleration fit}), so there would be a global
fit in both the transition trend and the acceleration floor which
predict the range of the minimal acceleration $a_{min}\sim(10^{-11}\sim10^{-10}m/s^{2})$. 

The dashed black curve is given by the transition trend (\ref{eq:interpolating MOND})
around $O(a_{0})$, i.e. 
\begin{equation}
a_{N}^{2}=\frac{a^{2}}{\sqrt{1+\frac{9}{4}\frac{a_{0}^{2}}{a^{2}}}}\label{eq:dash curve in fig}
\end{equation}
in which by $a_{0}=1.2\times10^{-10}m/s^{2}$ is used. The approximation
of the formula is hold around $O(a_{0})$ (dash-dot line) which is
a normalized version of $O(\sqrt{R_{0}})\sim O(\sqrt{\Lambda})$.
But we have continuously extended it to the unreachable deep-MOND
limit, which agrees to the fitting (\ref{eq:radial acceleration fit})
(red curve) within uncertainties. There is also problem in the predicted
dash curve in the Figure 2, although an acceleration floor is qualitatively
predicted in the deep-MOND limit, the acceleration floor predicted
at $a_{min}\sim10^{-10}m/s^{2}$ appears higher than the observational
hint if the transition trend is fit. As a result, the validity range
of the transition trend is smaller than the range of the observed
blue data points.
\begin{figure}
\begin{centering}
\includegraphics[scale=0.45]{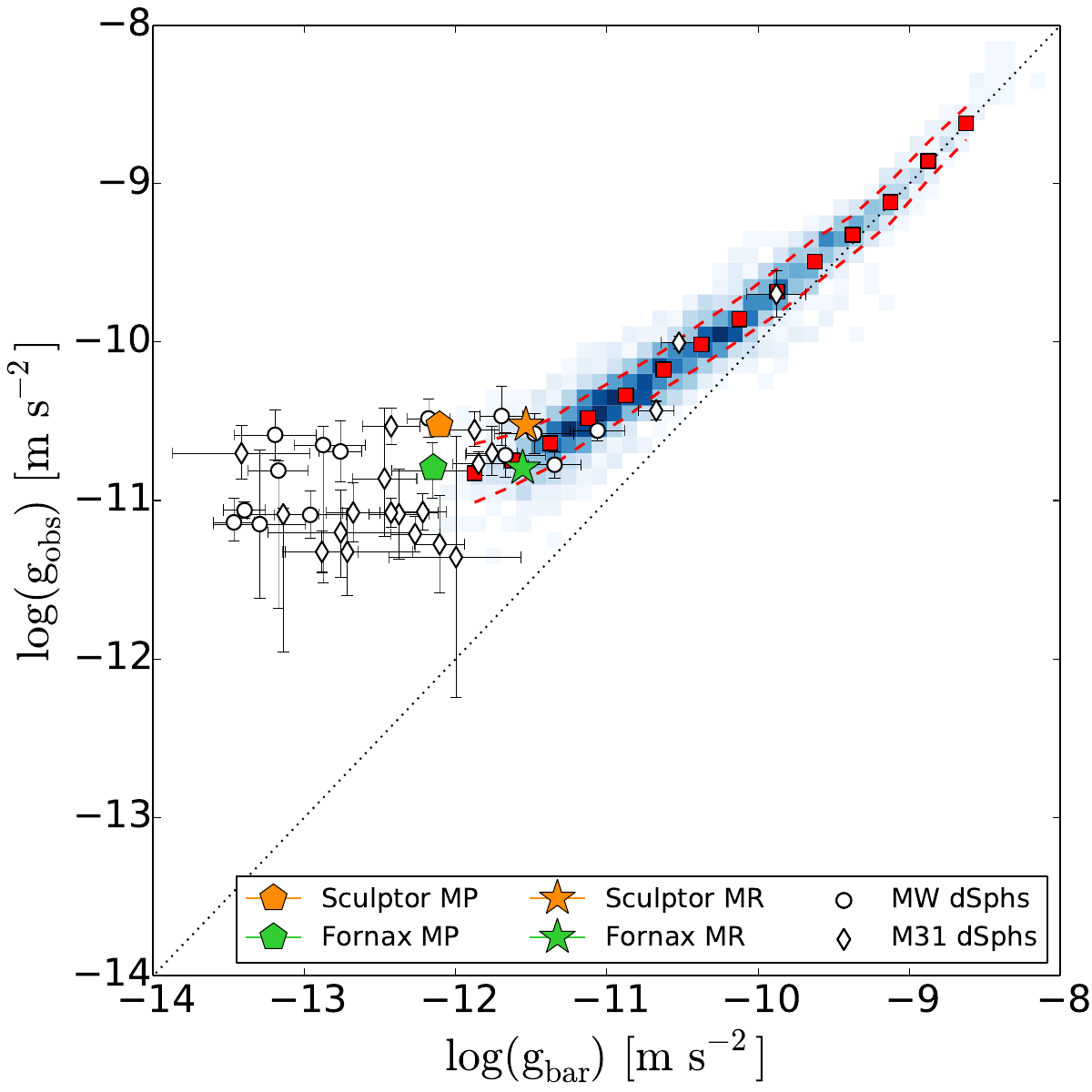}
\par\end{centering}
\caption{The empirical radial acceleration relation, citing from the Figure
11 of \citep{Federico2017One}. The blue square data points followed
by 2693 points in 153 galaxies, and the red point represent their
means. The circles (Milky Way dSphs) and diamonds (M31 dSphs) given
by the dwarf Spheroidals (dSphs) data points shows a possible ``acceleration
floor'', especially for those ultrafaint dSphs.}

\end{figure}
\begin{figure}
\begin{centering}
\includegraphics[scale=0.55]{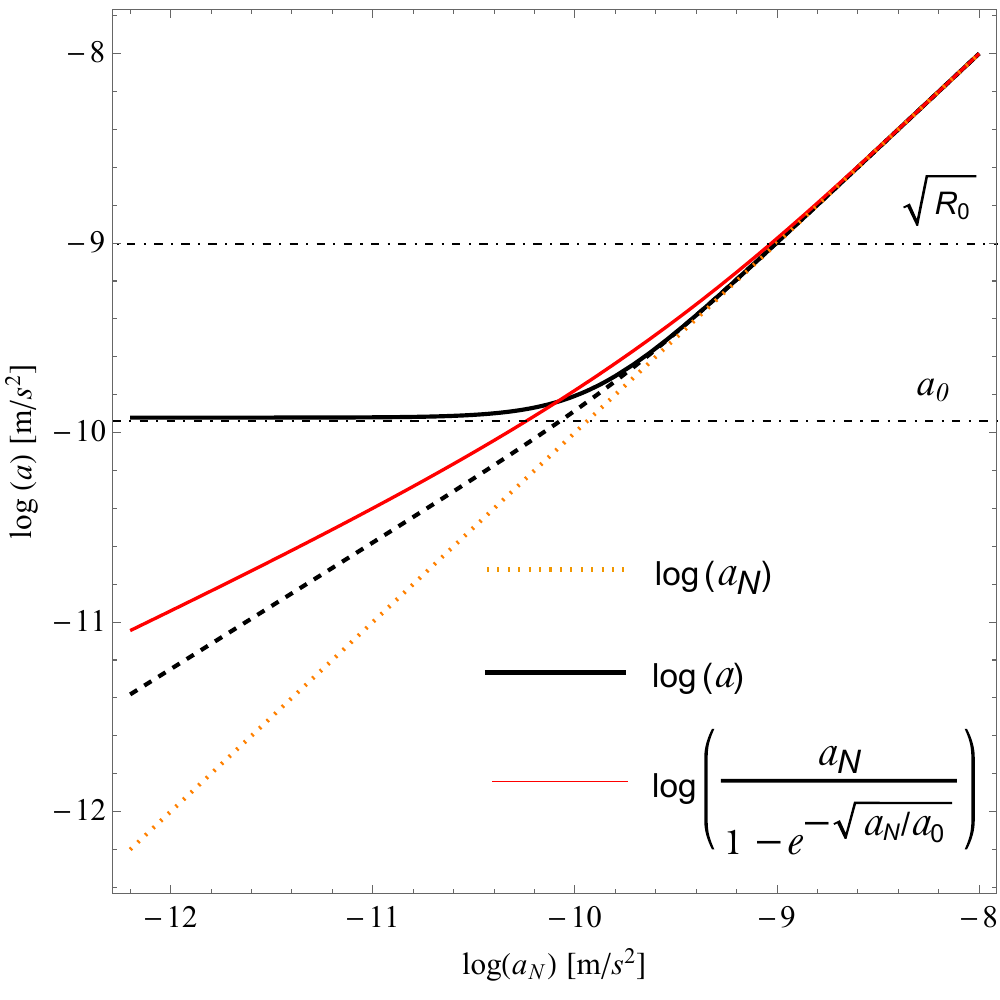}
\par\end{centering}
\caption{The radial acceleration discrepancy plots comparing with the Figure
1. The horizontal axis or the dotted orange line is the expected (Newtonian)
acceleration from the baryonic matter. The red curve is the eq.(\ref{eq:radial acceleration fit})
fitting the blue squares of the Figure 1, which has been continuously
extended to the unreachable deep-MOND limit beyond the data points.
The black curve is the acceleration $a$, assuming $a=\left(1-\frac{1}{2}R_{b}\Delta t\right)|\nabla\Phi_{g}|=\left(1-\frac{1}{2}R_{b}\Delta t\right)\sqrt{\frac{1}{2}R_{g}}$,
w.r.t. the expected (baryonic Newtonian) acceleration $a_{N}=\left(1-\frac{1}{2}R_{b}\Delta t\right)|\nabla\Phi_{N}|=\left(1-\frac{1}{2}R_{b}\Delta t\right)\sqrt{\frac{1}{2}\left(R_{g}-2\Lambda\right)}$,
so $a=\sqrt{a_{N}^{2}+\left(1-\frac{1}{2}R_{b}\Delta t\right)^{2}\Lambda}\approx\sqrt{a_{N}^{2}+a_{0}^{2}}$.
Assuming that the proper motion of test rotation satellites is relatively
small w.r.t. the expansion of the universe at high redshift, where
the background expansion acceleration of spacetime instead become
important, in this case the acceleration $a$ directly has an asymptotic
background acceleration floor corresponding to a universal cosmic
expanding acceleration at about $a_{min}\sim a_{0}\sim\left(1-\frac{1}{2}R_{b}\Delta t\right)\sqrt{\Lambda}\sim O(10^{-10}m/s^{2})$
shown in the Figure 2, where $\frac{1}{2}R_{b}\Delta t\sim\left(0.82\sim0.87\right)$.
If further considering a normalization to the acceleration as a curvature
square-root (see (\ref{eq:flow of R and g})) by $\left(1-\frac{1}{2}R_{b}\Delta t\right)^{1/2}$,
it is also plausible to further normalize $a_{min}$ to $a_{min}^{re}\sim\left(1-\frac{1}{2}R_{b}\Delta t\right)^{1/2}a_{0}\sim O(10^{-11}m/s^{2})$
in Figure 1. If the redshift of a galaxy is not too high, then the
proper rotation is still important w.r.t. the background expansion
at its outskirt. In this case, the acceleration floor may not directly
appear, and hence the dashed black curve can be continuously extended
along the transition trend of (\ref{eq:interpolating MOND}) to a
certain degree towards the unreachable asymptotic deep-MOND limit.
It has similar behavior with the fitting red curve within uncertainties.
And the acceleration floor is expected to asymptotically appear eventually
at some low acceleration region. Note that here $a_{0}$ plays the
role of a normalized version of $\sqrt{\Lambda}$ or $\sqrt{R_{0}}$,
so the approximation region (\ref{eq:coarse-graining approximation})
is hold around $O(a_{0})$. Comparing with the Figure 1, the acceleration
floor predicted at about $a_{min}\sim10^{-10}m/s^{2}$ appears higher
than the observational hint if the transition trend is fit. As a result,
the validity range of the transition trend is smaller than the range
of the observed blue data points.}
\end{figure}

\section{The ``Missing Matter'' Between Scales}

In order to have a natural interpretation of the required value $\frac{1}{2}R_{b}\Delta t\approx\left(0.82\sim0.87\right)$,
we consider the ``missing matter'' interpretation of the curvature
flow as the next order correction to gravity. The interpretation is
more or less similar with the renormalization effect of electric fields,
in which the ``polarization charge'' appears as the ``missing charge''
w.r.t. the ``free charge'' (visible charge) obeying the Maxwell's
equation. Now since the scalar curvature $R_{g}$ at the galactic
scale is larger than the $R_{b}$ at the baryonic or fiducial lab
scale, leading to the matter density at the galactic scale seems larger
than the visible baryonic matter, if one assumes the validity of the
classical Einstein's equation. Then some matters are seen ``missing''
in the galactic scale beside the visible baryonic matter and the ``dark
energy'' (the cosmological constant as the leading order correction).
When the baryonic matter is slowly moving and pressure-less, using
the classical Einstein equation $\Delta R=8\pi G\Delta\rho$, the
flow change of the scalar curvature (\ref{eq:scalar curvature flow})
can be translated to the equivalent ``missing matter'' density $\Delta\rho=\rho_{g}-\rho_{b}$,
\begin{equation}
\Delta R=R_{g}-R_{b}=R_{b}\left(\frac{1}{1-\frac{1}{2}R_{b}\Delta t}-1\right)=8\pi G\Delta\rho.
\end{equation}
Since the baryonic matter density gives $R_{b}=8\pi G\rho_{b}$, by
using the value of (\ref{eq:correction of a0}), we have
\begin{equation}
\frac{\Delta\rho}{\rho_{b}}=\frac{\frac{1}{2}R_{b}\Delta t}{1-\frac{1}{2}R_{b}\Delta t}\approx\left(4.6\sim6.7\right),
\end{equation}
which generally means the ``missing matter'' in the invisible halo
of galaxy at the galactic scale seems about $4.6\sim6.7$ times larger
than the baryonic matter in the luminary region of galaxy, roughly
consistent with the cosmological observations that the ``dark matter''
density is about 5 times the visible baryonic matter. And according
to the monotonicity of the Ricci flow, $\Delta R$ is non-negative,
so the ``missing matter'' $\Delta\rho$ is always non-negative as
well, which is also consistent with the observational fact that no
dark matter with negative mass. From this point of view, it is very
possible that the so called ``dark matter'' is just a mirage of
the Ricci flow of curvature. As long as the Ricci flow of spacetime
is real in physics, this kind of ``dark matter'' is inevitable,
at least contribute to a portion of the dark matter in the conventional
sense, even if not the whole.

We shall not detailed compare this theory with the dark matter theory
in the paper, since in fact we know little about the dark matter.
Here we could do some simple qualitative calculations for the ``missing
stress tensor $\Delta T_{\mu\nu}$'' and its Equations of State $w$
(EoS). The spatial component of the stress tensor is
\begin{equation}
\Delta T_{ij}=\frac{1}{8\pi G}\frac{\delta\left(\Delta R\right)}{\delta g^{ij}}=\frac{1}{8\pi G}\Delta R_{ij}.
\end{equation}
Obviously, in the non-relativistic galaxies, since matter at the initial
baryonic scale is non-relativistic, i.e. $R_{ij}(t_{b})\approx0$
as the initial condition of the Ricci flow, then we must have $\Delta R_{ij}\approx0$
as well during the Ricci flow, consequently the ``missing stress''
related to it gives $\Delta T_{ij}\approx0$ and EoS $w\approx0$.

In the background of Hubble expansion, we have $R_{\mu\nu}=\frac{R}{4}g_{\mu\nu}\approx3H_{0}^{2}g_{\mu\nu}$,
so
\[
\Delta T_{ij}\approx\frac{3H_{0}^{2}}{8\pi G}\Delta g_{ij}\approx O(\lambda)\delta_{ij}
\]
For its pressure is given by $\Delta T_{ij}=\Delta pg_{ij}=\Delta p\mathbf{a}^{2}\delta_{ij}$,
so we obtain a pressure proportional to the the scale factor $\mathbf{a}^{2}$
of the Hubble expansion. In a sufficiently redshifted galactic scale,
its pressure is sufficiently suppressed by the scale factor $g_{ij}=\mathbf{a}^{2}\delta_{ij}\ll\delta_{ij}$
at the high redshift and hence seems ``cold'' taking the equation
of state (EoS) $w\approx0$. Therefore, the ``missing matter'' and
the conventional dust-like baryonic matter are diluted in the same
way with the redshift, and hence the ratio $\frac{\Delta\rho}{\rho_{b}}\approx\left(4.6\sim6.7\right)$
is almost unchanged in the expanding history of the universe. While
at low redshift, the ``missing matter'' halo surrounding a galaxy
becomes warm but the pressure is still as low as $O(\lambda)\approx(10^{-3}\textrm{eV})^{4}$.

Certainly, the classical Einstein's equation is not exactly true in
the framework, so in fact the ``missing matter'' is not real matter,
it is just the mirage of the flow of curvature or gravity, in other
words, the ``missing matter'' is part of the gravity itself arisen
from the Ricci flow of initial spacetime from the baryonic matter.
Thus although the ``missing matter'' seems not possible pressure
support or rotation support, it still does not expected collapse gravitationally,
if the pure gravity itself is intrinsically non-collapsing \citep{Luo:2022goc},
so the induced the ``missing matter'' halo surrounding a galaxy
is considered stable.

In the framework, there are several observations in the internal relation
between the ``missing matter'' and the cosmological constant. First,
roughly speaking, the ``missing matter'' term $R\Delta t$ and $\nu\approx-\Omega_{\Lambda}$
could be of the same order shown in (\ref{eq:EH+cc}), as a consequence
that the ``dark matter'' and the ``dark energy'' (cosmological
constant) most naturally are both of the same order of the critical
density $O(\lambda)$ in the theory. Second, from (\ref{eq:nu}),
we can see in (\ref{eq:partition}) $\nu$ as a counter term measures
the whole difference between IR ($t=0$) and UV $t=-\infty$, so that
$\nu$ could completely cancel the scale change of the Shannon entropy
$\Delta\tilde{N}$ (higher order correction $O(R^{n}\tau^{n})$ might
appear in the Shannon entropy $\tilde{N}$ when beyond the low energy
expansion), leaving a short distance theory of the fiducial lab with
no ``missing matter'' and ``dark energy''. However, in galactic
observations, the ``missing matter'' term $\Delta R$ only measures
the finite scale difference, i.e. the difference between finite galactic
scale $t_{g}<0$ and baryonic scale $-\infty<t_{b}$, ($-\infty<t_{b}<t_{g}<0$),
so $-\nu$ do not completely cancel the ``missing matter'' $\Delta R$.
At galactic scale, it leaves a theory with mixture of $-\nu$ (dark
energy) and $R\Delta t$ (dark matter). Third, difference from $-\nu$
as a constant counter term, which leads to a ``4-spacetime volume
energy'' of EoS $w=-1$, the ``missing matter'' term $\Delta R$
is dynamic, its pressure can be suppressed by the expanding of the
3-space, so it is a ``3-spatial volume matter'' of EoS $w\approx0$,
thus they behave like different missing components of the universe. 

Further more, since $R\Delta t$ is roughly a constant of order one
shown previously, the fact might also imply that the distribution
of the supposed ``missing matter'' and corresponding curvature $R_{ij}(t_{g})$
(in the invisible halo surrounding a galaxy) can be modeled by the
local Shrinking Ricci Soliton equation (\ref{eq:gsrs}), $R_{ij}=\frac{1}{2\Delta t}g_{ij}$
(if simply setting a rough constant $u$ density in the galaxy). In
other words, the supposed ``dark halo'' may be nothing but a Shrinking
Ricci Soliton Ricci flow limit configuration seen by a long distant
observer. Since the Ricci flow tends to gradually smooth out the initial
inhomogeneous and anisotropic baryonic matter distribution, for instance,
an anisotropic disk galaxy. During the process, such local initial
configuration becomes more and more homogeneous and isotropic, thus
the ``dark halo'' of the disk galaxy will be rounder and rounder.
More precisely, the static Shrinking Ricci Soliton equation in 3-space
$R_{ij}=\frac{1}{2\Delta t}g_{ij}$ at linear approximation gives
a Helmholtz equation for the metric $\left(\varDelta+\frac{1}{\Delta t}\right)g_{ij}\approx0$
($\varDelta$ is the Laplacian operator of the space, $\Delta t$
the finite scale difference). The equation indicates a Yukawa type
profile with a characteristic radius $\sqrt{\Delta t}\approx\sqrt{t_{*}-t_{b}}$
of the round ``dark halo'', which is much larger than the radius
of the visible galaxy. It is analogous to the phenomenon that the
screening effect of a medium tends to gradually smooth out the inhomogeneous
and anisotropic free charges distribution by dipolarizing the dielectric
medium surrounding the free charges. During the process, the observed
total charges (free charge + polarization charge) distribution becomes
more and more homogeneous and isotropic. If the Shrinking Ricci Soliton
configuration really resembles the ``dark halo'' of a galaxy, it
also supports the possible explanation that the correction (\ref{eq:correction of a0})
of $a_{0}$ should be almost a constant. 

To sum up, the cosmological constant as the counter term of the Ricci
flow modifies the gravity at the leading order, manifesting transition
trend similar with MOND, but the deep-MOND limit is unreachable in
the theory. The theory not only eliminates the leading missing mass
components in the universe but also gives a universal and leading
part of the critical acceleration $a_{0}=\frac{2}{3}\sqrt{2\Lambda}$,
while it does not remove the missing components completely and $a_{0}$
is still several times larger than the best fit, the next order contribution
is required. The Ricci flow induces the ``cold'' missing matter
and a correction term $\frac{1}{2}R_{b}\Delta t$ to $a_{0}$. In
fact, a galaxy spacetime configuration just resembles but exactly
be a Shrinking Ricci soliton, so the correction term $\frac{1}{2}R_{b}\Delta t$
is not precisely a constant. Then the next order correction depends
on specific scales of different cosmic objects, and in essential non-universal.
In this sense, the theory seems like a mixture of MOND and the ``cold
missing matter''. Considering the ``missing matter'' is about 5
times the baryonic matter, then we obtain a roughly universal baryonic
Tully-Fisher relation of spiral galaxies with the best fit $a_{0}=\frac{2}{3}\sqrt{2\Lambda}\left(1-\frac{1}{2}R_{b}\Delta t\right)\approx\frac{\sqrt{\Lambda}}{\left(6\sim8\right)}$.
The correction term $R_{b}\Delta t$ here not only gives a correction
to $a_{0}$ but also the density of ``missing matter''. At this
point, it is treated as an approximate constant correction, if we
consider the galactic scale $t_{g}$ is close to the finite singular
scale $t_{g}\approx t_{*}$. But in fact $t_{g}$ is not precisely
$t_{*}$, it is yet unclear whether it is feasible to have a more
detailed testing of the scale $t_{g}$ or curvature dependence for
different galaxies. 

\section{Discussions and Conclusions}

In the paper, starting from the Ricci flow of quantum spacetime reference
frame as the first principle, a quantum modified gravity is obtained
at low energy and long distance scale via the small $\tau$ scale
parameter expansion. As the gravity theory has a low characteristic
energy scale $\lambda$, the gravity suffers from a major correction
at low energy. The Ricci flow effects to the modification of gravity
is two folds: the effect of the counter term of the Ricci flow and
the Ricci flow of the specific spacetime metric and curvature. The
low energy effective action is rather conservative, it can be simply
considered as a Ricci flowing Einstein-Hilbert action plus a cosmological
constant. 

The leading correction to the standard Einstein's gravity is a cosmological
constant as a counter term of the Ricci flow, which makes the spacetime
asymptotic a Shrinking Ricci soliton or deSitter. It not only modifies
the behavior of gravity at the cosmic scale (i.e. acceleration expansion
of the universe), but also contributes to the acceleration discrepancies
at the galactic scale (i.e. the baryonic Tully-Fisher relation, and
the radial acceleration discrepancies of galaxies) when the local
curvature is low enough to be comparable with the characteristic background
curvature $R_{0}$ corresponding to the low energy scale $\lambda$.
The leading low energy correction of the gravity coming from the cosmological
constant gives a similar transition trend and radial acceleration
discrepancy like MOND. This suggests a simple picture that the asymptotic
behavior of acceleration discrepancy around $a\sim(10^{-11}\sim10^{-9})m/s^{2}$
or $O(a_{0})$ can qualitatively come from the local effect of the
cosmological constant. The internal relation between the cosmological
constant and the transition trend of MOND is manifested in such a
way that the global effects of the cosmological constant or asymptotic
background curvature affects a local gravitational system, such as
galaxy, around the curvature $O(R_{0})$ and around the related acceleration
$O(a_{0})$.

However, the cosmological constant seems prevent the observed acceleration
being much lower than a coarse-grained lower bound, so it gives a
flattening ``acceleration floor'' when $a_{N}\ll a_{0}$ or $\rho_{M}\ll\lambda$
at the outskirt of galaxies, which is qualitatively consistent with
some hints of the observations. When some correction factor is introduced,
the predicted bound could roughly close to the observed bound. The
existence of the ``acceleration floor'' differs from MOND at the
deep-MOND limit $a_{N}\ll a_{0}$. If the acceleration is able to
reach the deep-MOND limit, then as $a\rightarrow0$, the final constant
rotation curve $v^{2}=a\cdot r$ of a galaxy would be extended to
infinity $r\rightarrow\infty$, which is impossible from the cosmological
viewpoint. It is more natural to exist an escaping maximal radius
of rotating satellites beyond which the curvature and related acceleration
achieve minimum, rather than reaches the deep-MOND limit $a\rightarrow0$.
In fact the deep-MOND limit might not be applied the principle of
general covariance \citep{Milgrom:2019rtd}. In general, since in
the framework, the acceleration (defined by the second spacetime derivative
of the coordinate) of a test satellite (spectral line) at the outskirts
of a galaxy is affected by the variance or second moment fluctuation
of the spacetime coordinate, if the Ricci flow and the second moment
quantum fluctuation of spacetime coordinate at long distance scale
are intrinsically unavoidable, and the rotation velocity and related
acceleration of the satellites are essentially measured by the broadening
(variance) of the spectral lines, there would very likely be a final
spectral line broadening due to the Ricci flow and a related fundamental
lower bound for the acceleration of the test satellites. In the framework,
the acceleration discrepancy observed at the outskirts of galaxies
arises from a process that is initially dominated by Doppler broadening
caused by local proper motion (baryonic matter giving), but gradually
becomes dominated by quantum broadening caused by background spacetime
expansion (Ricci flow giving). From this point of view, if we desire
a unified view of gravity that not only modifies its behavior at the
outskirts of a galaxy but also accurately accounts for the cosmological
scale, a deviation from the deep-MOND behavior at the cosmic scale
seems inevitable. 

The Ricci flow of metric or curvature plays a role in providing next
order corrections to gravity at low energy. The starting scale used
to calibrate measurements by optical law and to construct the distance
ladder is different from the observed scale, such as the cosmic or
galactic scale, which is relatively redshifted and quantum broadened.
The scale change due to the Ricci flow as a next order correction
is non-negligible. The first consequence of the flow effect is that
it gives a scale dependent correction to the universal part of the
critical acceleration $a_{0}$, which can be interpreted as the normalization
of the coupling between the matter and the potential or acceleration.
The corrected $a_{0}$ can be consistent with the observed fitting
value for galaxies, if the equivalent \textquotedblleft missing matter\textquotedblright{}
in the invisible halo of galaxy is about five times the baryonic matter
in the luminary region of galaxy. The second consequence of the flow
effect is that it is equivalent to the \textquotedblleft missing mass\textquotedblright{}
mimicking the cold dark matter with the equation of state $w\approx0$
at high redshift, while at low redshift the \textquotedblleft missing
matter\textquotedblright{} becomes warm with low enough pressure.

In the framework, if we consider the leading order correction to gravity
from the cosmological constant is intrinsically universal, the next
order are corrections depending on specific scales $t_{g}$ of various
cosmic objects, only approximately universal if galaxy spacetime configuration
resembles a Shrinking Ricci soliton limit. In this picture, a rough
distribution of the ``missing matter'' invisible halo surrounding
a visible galaxy is approximately determined by a Shrinking Ricci
soliton equation, $R_{ij}=\frac{1}{2\Delta t}g_{ij}$. The Ricci flow
smooths out the initial inhomogeneous and anisotropic baryonic matter
distribution of a local visible galaxy, making the local spacetime
more and more like a local homogeneous and isotropic Shrinking Ricci
soliton configuration as the flow limit, and making the ``missing
matter'' halo surrounding the galaxy rounder and rounder as the coarse-graining
process of the Ricci flow tends to do. This picture might provide
us a possible way to model the profile of an invisible halo of galaxy
by the Ricci flow approach. 

The leading order and next order corrections to gravity at low energy
produce a theory like a mixture of MOND and \textquotedblleft cold
missing matter\textquotedblright , similar to the $\Lambda$CDM model,
which is mainly a mixture of the cosmological constant and cold dark
matter. The difference is that the transition trend and \textquotedblleft cold
missing matter\textquotedblright{} here are correlated to each other
and have a common origin: the Ricci flow of spacetime. To understand
the astronomical phenomenon, both of them are necessary. The framework
is not motivated by empirical fitting of astronomical data, but purely
a framework of gravity coming from alternative theoretical considerations,
which happened to contain these speculative \textquotedblleft dark\textquotedblright{}
components. The Ricci flow might provide us with a different paradigm
for cosmological and astronomical observations.

The acceleration expansion of universe and acceleration discrepancy
of galaxies given by the low energy quantum modified gravity has a
unified understanding by the fact that the quantum spacetime with
2nd order moment fluctuation modifies the physical quantities at 2nd
order in spacetime coordinates, like the curvature and acceleration,
which are second spacetime derivative induced. In other words, curvature
and acceleration can be roughly seen as the same thing in the sense
that they both suffer from the quantum fluctuation corrections at
the same (2nd) order. In the framework, the curvature ratio $R/R_{0}$
plays a similar role of the acceleration ratio $a/a_{0}$ in MOND,
but essentially, the curvature ratio is more fundamental and has many
advantages than the acceleration ratio. Curvature is a more fundamental
and general covariant concept, and hence the framework is a relativistic
theory. But the concept of acceleration is not generally covariant,
which might lead to confusions in MOND, e.g. the acceleration is absolute
or relative to any specific frames, or direction dependent, internal
or external, etc.? Some relativistic generalizations of MOND could
be seen as attempts to solve this kind of conceptual difficulties
making it more well-defined especially in cosmology. The framework
of the paper is based on the first principle for alternative motivations,
and hence it differs from most of the relativistic generalization
of MOND. As one of the implications of MOND and its relativistic generalization,
the Equivalence Principle is considered explicitly violated. In fact
any modifications of gravity has to face the issue that the effective
gravitational mass differs from the inertial mass of the unmodified
gravity, or effective metric differs from the one of the unmodified
gravity, so it seems that one has to deal with at least two spacetime,
but only one of which we can sense directly, e.g. the bi-metric theory
\citep{Milgrom:2009gv}. In some theories, the difference of masses
or metric of spacetime is interpreted as the violation of the Equivalence
Principle \citep{Smolin:2017kkb}. Or in some other theories, the
difference is alternatively interpreted as missing mass or other kind
of new matter e.g. cold dark matter and dipoler dark matter \citep{Blanchet:2009zu},
etc. However, it is worth stressing that the Equivalence Principle
retains in the framework even at the quantum level, the price to pay
is that the metric or curvature of the spacetime must flow or renormalize
with scale at the quantum level. Different metrics of spacetime (e.g.
$g(t_{g})$ and $g(t_{b})$) in the modified gravity at different
scales have equal realities. Since the Equivalence Principle plays
a fundamental role in the quantum reference frame theory. The Equivalence
Principle is the physical foundation of measuring the spacetime by
physical material reference frame even at the quantum level, and it
is the physical foundation of the geometric interpretation of gravity
through curved spacetime, so that the gravity is simply a relational
phenomenon that the motion of a test particle in gravity is manifested
as a relative motion w.r.t. the (quantum) material reference frame.
Without the Equivalence Principle, we would lost the physical foundation
of all these concepts, such as the metric and acceleration that all
the theories are based on.

There are limitations in the studies. The transition trend away from
Newtonian gravity of the theory is studied and tested with reference
to the universal and successful part of MOND. However, the theory
and MOND have completely different behaviors at the deep-MOND limit
$a_{N}\ll a_{0}$. In this picture, although an acceleration floor
is qualitatively predicted in the deep-MOND limit, the acceleration
floor predicted at about $a_{min}\sim10^{-10}m/s^{2}$ appears higher
than the observational hint if the transition trend is fit. As a result,
the validity range of the transition trend is smaller than the range
of the observed blue data points. Thus, if the idea that the MOND-like
transition trend and the acceleration floor are both due to the cosmological
constant (i.e. its broadening effect to the spectral lines) is on
the right track, further effects must be taken into account. The question
of the acceleration floor is still open in the paper. It is also unclear
whether the full relativistic theory can solve some of the challenges
of MOND, such as the bullet cluster and/or missing baryons in groups
or clusters of galaxies. Qualitatively, the framework may provide
a promising picture, as clusters colliding or for other possible reasons,
it is possible to observe a separation of the mass-centers of the
baryonic matter and the effective \textquotedblleft missing matter\textquotedblright{}
in a full relativistic theory. This is analogous to the separation
of the charge-centers of the free charge and the polarization charge
when the free charge is sharply accelerated, and the response of the
polarization charge is delayed due to its own inertia. Additionally,
it is unclear whether the full scale-dependent theory can satisfactorily
explain some of the sharp challenges in cosmology that are considered
to be the big successes of CDM, such as a consistent history of structure
formation and fitting the third and subsequent peaks of the acoustic
spectrum in the Cosmic Microwave Background. Furthermore, there are
several unexplained discrepancies to the global fit $a_{0}$ reported
in e.g. \citep{Rodrigues:2018duc,10.1093/mnras/stu100}, but whether
these discrepancies are truly correlated to the curvature or scale-dependent
correction (\ref{eq:scale dependent a0}) is yet unclear. These issues
require further study.
\begin{acknowledgments}
This work was supported in part by the National Science Foundation
of China (NSFC) under Grant No.11205149, and the Scientific Research
Foundation of Jiangsu University for Young Scholars under Grant No.15JDG153.

\bibliographystyle{plain}

\end{acknowledgments}

\end{document}